# Concentration-Dependent Tungsten Effects on Chemical Short-Range Order and Deformation Behavior in Ni–W alloys

Shaozun Liu[1,6*], Zehao Li[2], Hantong Chen[3], Xingyuan San[4], Bi-Cheng Zhou[3*], Dieter Isheim[2,7], Tiejun Wang[1], Nie Zhao[5], Yu Liu[1], Yong Gan[1], Xiaobing Hu[2,8*]

[1]Central Iron and Steel Research Institute, Beijing 100081, China

[2]Department of Materials Science and Engineering, Northwestern University, Evanston, IL 60208, USA

[3]Department of Materials Science and Engineering, University of Virginia, Charlottesville, Virginia 22904, USA

[4]Hebei Key Lab of Optic-electronic Information and Materials, The College of Physics Science and Technology, Hebei University, Baoding 071002, China

[5]School of Materials Science and Engineering, Xiangtan University, Xiangtan 411105, China

[6]Yangjiang Advanced Alloys Laboratory, Yangjiang 529500, China

[7]Northwestern University Center for Atom Probe Tomography (NUCAPT), Evanston, IL 60208, USA

[8]The NU*ANCE* Center, Northwestern University, Evanston, IL 60208, USA

[*]Corresponding authors. Email: xbhu@northwestern.edu (X.H.); liushaozun@cisri.com (S.L.); bicheng.zhou@virginia.edu (B.C.Z.);

## Abstract

Ni–W based medium heavy alloys offer a promising pathway to bridge the density–strength gap between tungsten heavy alloys and ultrahigh-strength steels. In this study, the effects of W concentration on chemical short-range order (CSRO), deformation behavior, and grain boundary chemistry of Ni–*x*W in the range of *x*=0 to 38 wt.% alloys were systematically investigated using a suite of advanced characterization and modeling techniques, including synchrotron X-ray diffraction, transmission electron microscopy, atom probe tomography, and first-principles thermodynamic simulations. Our study reveals that strong CSRO emerges when W content exceeds ~30 wt.%, producing distinct diffuse scattering and significantly enhancing strain-hardening capacity. During deformation, the presence of CSRO promotes planar slip and twin formation, leading to strong dislocation interactions and elevated flow stress. Hall–Petch analysis demonstrates an exceptionally high grain boundary strengthening coefficient ($k_y \approx 1102$ MPa·μm$^{-1/2}$) in Ni–38W, underscoring the intrinsic strengthening effect associated with CSRO. First-principles cluster expansion coupled with Monte Carlo simulations reveals that increasing W content enhances CSRO tendency through the stabilization of $Ni_4W$ type local configurations. These findings establish a mechanistic link between W concentration, CSRO evolution, and mechanical response, providing new insights for designing high-density, high-strength Ni–W based alloys with optimized performance.

## 1. Introduction

Tungsten heavy alloys (WHAs) have long been recognized as indispensable structural and functional materials due to their high density, balanced strength and toughness, excellent electrical and thermal conductivity, and superior resistance to wear, corrosion, and oxidation [1-6]. These properties make WHAs vital in a wide range of applications, spanning from defense and aerospace to medical and nuclear industries. Their combination of mass and durability enables their use in armor-piercing projectiles, counterweights for missiles and aircraft, radiation shielding, and advanced electrical and mechanical tooling. Despite these advantages, the broad application of WHAs has been constrained by their limited mechanical performance, especially when benchmarked against ultrahigh strength steels (UHSSs) [7-13]. The root of this limitation lies in the intrinsic characteristics of the body-centered cubic (BCC) tungsten (W) matrix, which suffers from poor ductility and a restricted ability to accommodate plastic deformation. Consequently, while WHAs provide density advantages, they fall short in mechanical robustness compared with steel-based systems. By contrast, UHSSs exhibit exceptional strength and toughness, making them attractive for structural applications in automotive, aerospace, and energy sectors. However, their low density (~7.8–8.1 g/cm$^3$) is less than half that of WHAs (~16.5–18.5 g/cm$^3$), restricting their use in applications that demand high inertia, penetration, or shielding. The stark density-mechanical property trade-off between WHAs and UHSSs motivates the exploration of new materials that can bridge this performance gap.

Recently, Ni–W based medium heavy alloys (MHAs) have emerged as promising candidates to fill this niche. By combining the high density of W with the excellent deformability of face-centered cubic (FCC) nickel (Ni), MHAs exhibit a unique balance of density and mechanical performance. Our previously developed protype Ni-W based MHA with a composition of 57Ni-37W-5Co-1Ta (in wt%) demonstrates this balance effectively [14]. With the Ni-W solid solution-based FCC matrix, the protype MHA (~11.39 g/cm$^3$) achieves higher density than UHSSs while offering superior mechanical performance compared with traditional UHSSs [15, 16]. Unlike BCC-structured WHAs, which intrinsically suffer from limited deformability, our prototype and subsequently developed MHAs features a single-phase FCC Ni-W solid-solution matrix [14, 17]. This FCC matrix not only improves intrinsic deformability but also enables additional potential strengthening through deformation-induced mechanisms and potential secondary-phase precipitation. This distinctive microstructural framework highlights Ni–W based MHAs as a new generation of alloys with significant potential in defense, aerospace, and energy-intensive industries.

An intriguing feature of Ni-W based alloys, as revealed in our prior studies, is the presence of CSRO, which was observed in the forged condition and evolved into ordered $Ni_4W$ precipitates upon post-aging treatment [14]. CSRO is a fundamental structural phenomenon in solid solutions, characterized by non-random atomic arrangements over a few interatomic distance, which can profoundly influence the materials' thermodynamic stability [18], stacking fault energy [19, 20], irradiation response [21, 22], corrosion resistance [23], and mechanical behavior [24, 25]. In FCC alloys, CSRO has been reported to alter dislocation behavior, enhance solution strengthening, and in some cases act as a precursor to long-range ordered phases [26-28]. However, despite its significance, the understanding of CSRO in Ni–W based MHAs remains incomplete. In particular, the role of W concentration in governing the degree of CSRO, as well as its correlation with strengthening mechanisms and overall alloy performance, has not yet been systematically clarified. This knowledge gap is particularly important in the context of alloy design. For MHAs to evolve into a technologically viable class of materials, it is critical to establish a mechanistic understanding of the factors controlling CSRO formation and stability. Concentration-dependent CSRO behavior is especially relevant because the solute–solute interactions, lattice strain, and chemical bonding introduced by W atoms directly affect both the local ordering tendency and the resulting solution strengthening. A deeper comprehension of these effects will not only enable the rational design of Ni–W based alloys with optimized performance but also inform broader alloy development strategies across FCC-based medium- and high-density systems.

In this work, we aim to systematically investigate the influence of W concentration on CSRO formation, solution strengthening, and microstructural evolution in Ni–W alloys upon deformation. By combining advanced characterization techniques, including synchrotron X-ray diffraction, transmission electron microscopy (TEM), atom probe tomography (APT), and advanced theoretical calculations—with systematic evaluations of mechanical performance, we reveal the intrinsic role of W in governing CSRO behavior and the consequent effects on mechanical performance. The correlation among W concentration, CSRO, and strengthening response is established. The insights gained not only advance the fundamental understanding of CSRO in Ni–W systems but also provide a knowledge base for the design of next-generation MHAs that bridge the density and performance gaps between WHAs and UHSSs.

## 2. Experimental procedures and theoretical calculations

The Ni-W binary alloys with varying W contents were prepared using the vacuum induction melting method. The nominal W contents were 0% (0%), 10% (3.44%), 20% (7.48%), 30% (12.16%), 34% (14.3%), and 38% (16.24%) in wt.% (in at. %), designated as Ni, Ni-10W, Ni-20W, Ni-30W, Ni-34W, and Ni-38W, respectively. After melting, the alloys were cast into ingots, reheated to 1200°C, and subjected to multiple forging passes to produce rods with a diameter of 20 mm. The forged alloys were subsequently solution-treated at 850 °C, 900 °C, 950 °C, 1000 °C, 1050 °C, and 1100 °C for 1h, followed by water quenching. For optical microscopy observations, samples were mechanically grounded, polished, and etched using a solution containing 2 g $CuCl_2$, 40 mL HCl, and 60 mL ethanol. Tensile tests were carried out at room temperature on the INSTRON 5582 testing machine at a strain rate of $5\times10^{-4}$ $s^{-1}$. The tensile specimens had a standard geometry with a gauge section diameter of 5 mm, and a gauge length of 25 mm. X-ray diffraction (XRD) analyses were performed using a Bruker D8 diffractometer with Cu radiation (λ=1.541 A˚). Synchrotron radiation XRD measurements were conducted at the BL13SSW beamline of the Shanghai Synchrotron Radiation Facility using the photon energy of 50 keV (λ=0.248 A˚) and 97.9 keV (λ=0.127A˚). TEM characterizations were carried out on the ARM 200CF microscope equipped with a probe lens corrector. The microscope was operated at 200 kV. Electron transparent thin foils were prepared by conventional ion milling. Three-dimensional APT was employed to investigate the grain boundary chemistry of Ni–W alloys using a CAMECA local electrode atom probe (LEAP) 5000XS instrument with a detection efficiency of approximately 80%. The APT datasets were collected in laser pulsing mode at a base temperature of 50 K, with a laser pulse energy of 30 pJ and a pulse frequency of 250~500 kHz. Three-dimensional reconstructions and data analysis were processed with the AP Suite 6.3 software package. Needle-shaped specimens for APT analysis were prepared using a Thermo Fisher Helios 5 plasma focused ion beam (FIB) system equipped with a Xe source and electron backscatter diffraction (EBSD) detector. To identify the grain boundary (GB) located near the specimen's apex and misorientation angles across GBs, transmission Kikuchi diffraction (TKD) was also performed after FIB sharpening using an Oxford Symmetry S3 detector with an accelerating voltage of 25 kV, probe current of 1.6 nA, and step size of 5 nm.

First-principles thermodynamic calculations were performed to investigate the origin and characteristics of CSRO in Ni-W alloys. Cluster expansion (CE) coupled with Monte Carlo (MC) simulations were performed to predict the CSRO tendency of the system. In this work, we used a recently developed Bayesian CE method implemented in the Alloy Theoretical Automatic Toolkit (ATAT) [29-32]. This approach ensures an exact fit for all ground states by construction while significantly reducing the computational cost. Since FCC W is mechanically unstable, the inflection detection method was applied to obtain the precise reference energy [33-35]. To train the CE model, 30 structures near the Ni-rich end of the Ni-W system were selected as references configurations. First-principles calculations were carried out using the Vienna *ab initio* Simulation Package (VASP) [36-39], employing the projector augmented wave (PAW) method and the Perdew–Burke–Ernzerhof (PBE) exchange-correlation functional, with an energy cutoff of 520 eV [39-41]. Details regarding the effective cluster interaction (ECI) parameters and the clusters used in the CE fitting are available in the GitHub repository via the link provided in the supplementary materials. K-point meshes were automatically generated through ATAT to ensure a minimum density of 2000 k-points per reciprocal atoms [29].

Following the CE construction, MC simulation was conducted to determine the phase boundaries and CSRO evolution up to 20 at% of W. The *phb* module in ATAT [29] was first used to predict the two-phase region separating FCC Ni and $Ni_4W$. Equilibrium atomic correlations were then obtained using the *memc2* code [29], and the corresponding cluster concentrations were derived using the V-matrix formalism [42], which was implemented in the *cvmclus* module in ATAT [43]. To quantify CSRO in the Ni-W system, the Warren-Cowley CSRO parameter was calculated as [44]:

$$\alpha_{ij} = 1 - \frac{P_{ij}}{2c_i c_j} \qquad (1)$$

where $\alpha_{ij}$ is the CSRO parameter for the nearest neighbor pair *i-j*, $P_{ij}$ is the probability of finding such a pair among all first nearest neighbor pairs, and $c_i$ and $c_j$ are the atomic fractions of elements $i$ and $j$, respectively. The CSRO parameters are computed for various temperature-composition conditions using MC simulations, and the resulting Warren-Cowley CSRO heat map was superimposed on the conventional phase diagram to construct the CSRO phase diagram of the Ni-W system.

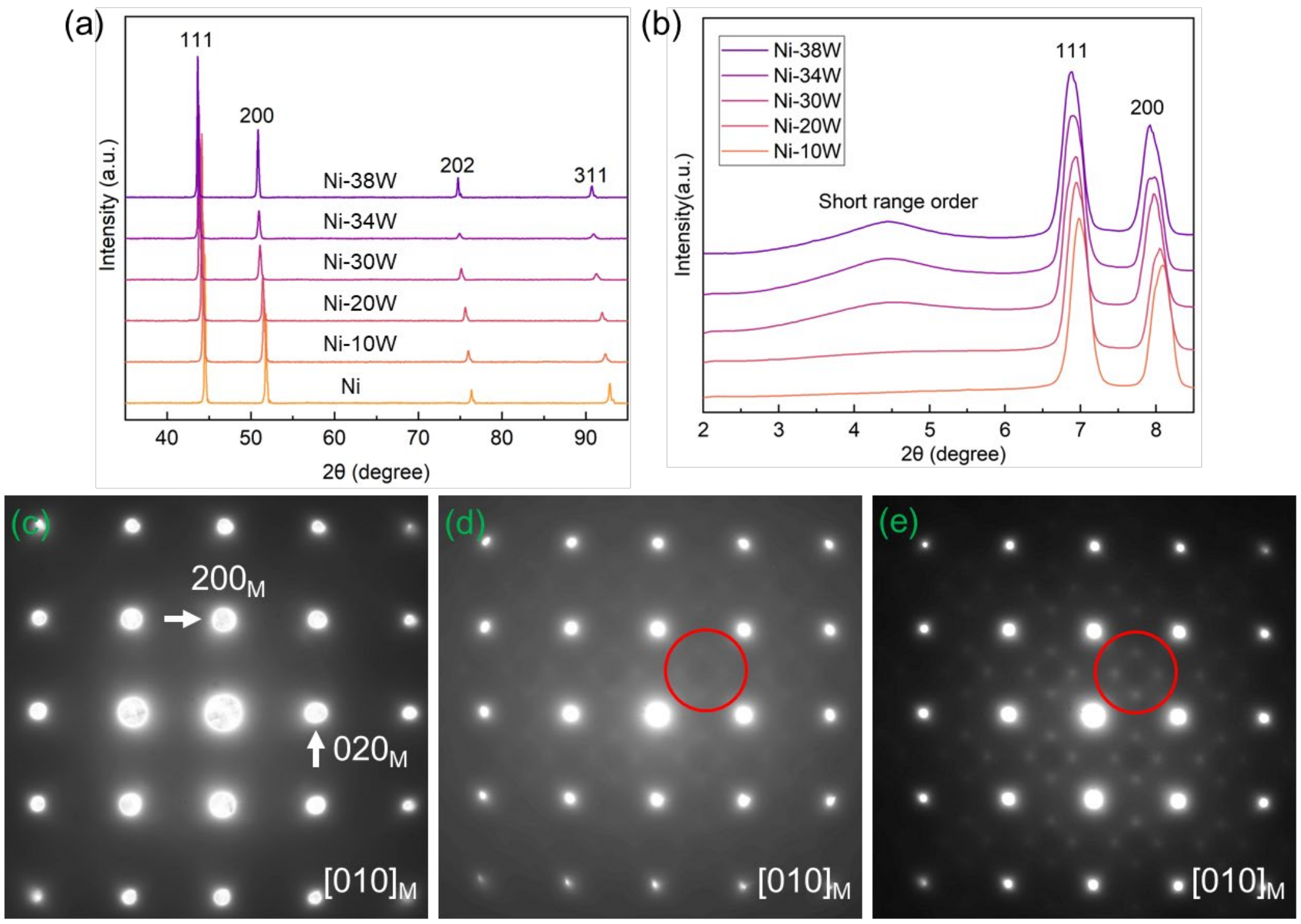

**Figure 1.** (a) Traditional XRD results of various Ni-W alloys. (b) Low angle synchrotron radiation XRD results of Ni-W alloys with varied W contents. The used photon energy here is 50 keV. Selected area electron diffraction (SAED) patterns of (c) Ni-10W, (d) Ni-30W, (e) Ni-34W samples. The arrows in (d, e) indicate the satellite patterns. All alloys investigated here were solution-treated at 1050°C for 1h.

## 3. Results

### 3.1 Concentration-dependent CSRO behavior

Across all investigated W contents, the solution-treated Ni-W alloys exhibit a single-phase FCC structure, as confirmed by XRD patterns in **Fig. 1a**. Synchrotron radiation XRD data (**Fig. 1b**) reveal broad diffusion peaks around 4.5° in alloys with W concentration exceeding 30%. The intensity of these diffuse peaks increases with higher W concentration, indicating the progressive development of CSRO. The SAED patterns in **Figs. 1c-e** correspond to Ni-10W, Ni-30W and Ni-34W, respectively, all of which can be indexed along the $[001]_M$ zone-axis, where the subscript M denotes the FCC Ni-W matrix. Compared to **Fig. 1c**, distinct satellite patterns at $\left\{\frac{1}{4}\frac{1}{2}0\right\}_M$ are observed in **Figs. 1d** and **1e**, confirming the formation of CSRO within the solid-solution matrix [14, 45]. Furthermore, the satellite reflections in Ni-34W are noticeably stronger than those in Ni-30W, consistent with the bulk XRD observations.

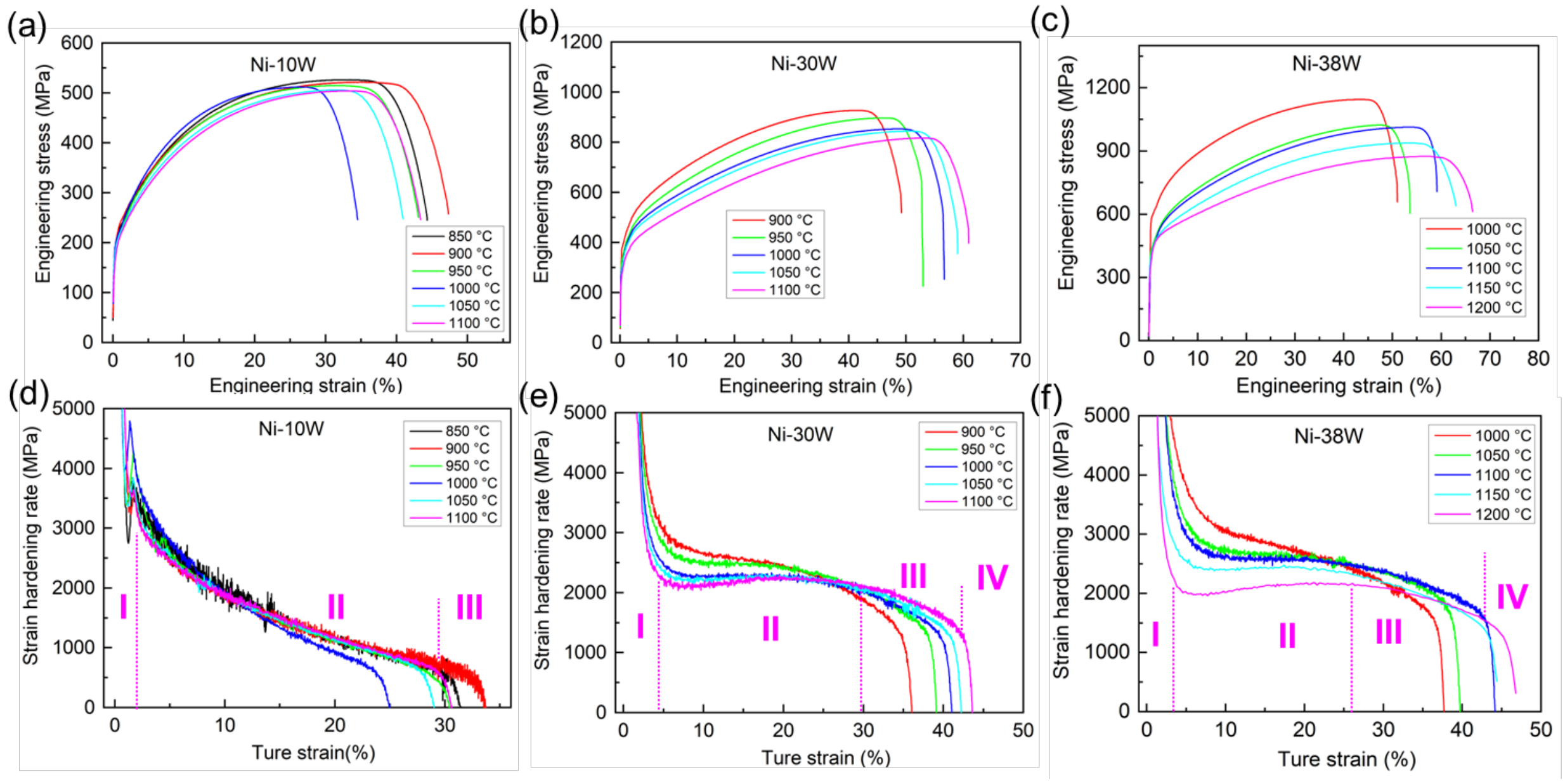

**Figure 2.** Engineering stress-strain and strain hardening rate curves of Ni-W alloys at different solution temperatures. (a, d) Ni-10W, (b, e) Ni-30W, (c, f) Ni-38W. The solution temperature for each curve is indicated in the legend. The vertical pink dash lines in (d-f) highlight the distinct deformation stages for Ni-10W solution-treated at 1100 °C, Ni-30W solution-treated at 1100 °C and Ni-38W solution-treated at 1200 °C respectively.

### 3.2 Concentration-dependent mechanical behavior

To elucidate the effect of W on grain-refinement strengthening, a series of solution treatments at different temperatures were performed on Ni-W alloys with varying W contents, followed by optical microscopy characterizations (Supplementary **Figs. S1-S5**). The results indicate that the recrystallization temperature increases with W concentration. Alloys containing less than 30 wt.% W were fully recrystallized at 900 °C, whereas Ni-34W and Ni-38W required higher temperatures of 950 °C and 1000 °C, respectively. Grain growth occurred at temperatures above the recrystallization threshold, with higher-W alloys exhibiting slower grain coarsening kinetics. The measured grain sizes for each alloy are summarized in **Table S1**. Meanwhile, it is noted that following comparisons of all deformation behavior and Hall–Petch relationships were performed using recrystallized alloys with comparable grain structures, minimizing the influence of incomplete recrystallization or heterogeneous microstructures.

Tensile stress-strain curves and strain-hardening rate plots for various alloys subjected to different solution-treatment temperatures, are presented in **Fig. 2** and Supplementary **Fig. S6**. The variations of yield strength (YS), ultimate tensile strength (UTS) and elongation of Ni-W alloys with different W concentrations and solution temperatures are summarized in **Table S2-S4**. Both YS and UTS increase markedly with W content. However, increasing the solution-treatment temperature leads to reductions in YS and UTS due to recrystallization and subsequent grain growth. **Fig. 2d** and **Fig. S6c** are the strain-hardening rate ($\theta = \frac{d\sigma}{d\epsilon}$) versus true strain curve of Ni–W alloys with lower W concentrations (<30 wt%) and negligible CSRO features. It is found that their deformations exhibit three characteristic strain-hardening stages. Taking Ni-10W treated at 1100 °C as an example (**Fig. 2d**), Stage I (0–2.5% strain) shows a rapid decrease in $\theta$ accompanied by a sharp increase in true stress. In Stage II (2.5–28% strain), the curve displays a quasi-linear relationship in which true stress continues to rise, albeit more gradually, while $\theta$ steadily declines. When the strain exceeds ~28%, Stage III begins, where $\theta$ decreases to zero or negative values promptly, signaling the onset of plastic instability and non-uniform deformation. In contrast, for alloys with higher W contents (≥30 wt%) and evident CSRO features, the strain-hardening behavior becomes distinctly four-staged (**Figs. S6d**, **2e**, **2f**). After the initial rapid decrease in $\theta$ during Stage I, a pronounced plateau appears and persists through most of the deformation process. In this regime, $\theta$ remains nearly constant or slowly declines, reflecting a dynamic balance between defects storage and recovery. For Ni-30W treated at 1100 °C (**Fig. 2e**), the plateau extends from approximately 5–30% strain (about 55% of the total deformation), while for Ni-38W heated at 1200 °C (**Fig. 2f**), it spans 4–27% strain (about 49% of total deformation). Beyond the plateau, both Ni-30W and Ni-38W exhibit dynamic recovery and softening like those in Ni-10W (**Fig. 2d**) and Ni-20W (**Fig. S6c**). This evolution highlights the critical role of CSRO in governing the deformation mechanisms and strain-hardening behavior of Ni–W alloys.

The mechanical behavior of Ni-W alloys is strongly governed by W concentration, solution-treatment temperature, and grain size. As shown in **Fig. 3a**, Young's modulus, density and lattice parameters all increase linearly with W content. The dominant strengthening mechanisms in these alloys are solid-solution and GB hardening, while contributions from precipitation and dislocation strengthening are negligible in recrystallized alloys. This is because deformation-induced defects introduced during forging are fully recovered, and no long-range ordered $Ni_4W$ precipitates are detected in the solution-treated alloys. GBs strengthening follows the Hall-Petch relationship: $\sigma_y = \sigma_0 + k_y d^{-\frac{1}{2}}$, where $\sigma_y$ is the yield strength, $\sigma_0$ is the friction stress for dislocation slip, $d$ is the grain size, and $k_y$ is the Hall-Petch coefficient quantifying the effect of grain size on yield strength [46-49]. As shown in **Fig. 3b**, YS scales linearly with $d^{-\frac{1}{2}}$ for all alloys, and both $\sigma_0$ and $k_y$ increase with W content. At 38% W, the Hall-Petch coefficient $k_y$ reaches an exceptionally high value of $\sim 1102\ \mathrm{MPa} \cdot \mu\mathrm{m}^{-\frac{1}{2}}$, exceeding nearly all previously reported values [50, 51]. For comparison, the Ni-Mo system exhibits a maximum $k_y$ of $\sim 1028\ \mathrm{MPa} \cdot \mu\mathrm{m}^{-\frac{1}{2}}$ at the Mo's solubility limit [50]. The relative evolution of $\sigma_0$ and $k_y$ with increasing W content is shown in **Fig. 3c**. The variations of both parameters can be divided into two distinct stages, with a transition occurring around 30 wt.% W — corresponding to the threshold for significant CSRO formation. Beyond this transition, the increase in $k_y$ becomes markedly more pronounced, while the growth rate of $\sigma_0$ slows compared to the initial stage without CSRO. This two-stage behavior indicates that CSRO plays a crucial role in governing the deformation mechanisms of Ni-W alloys process.

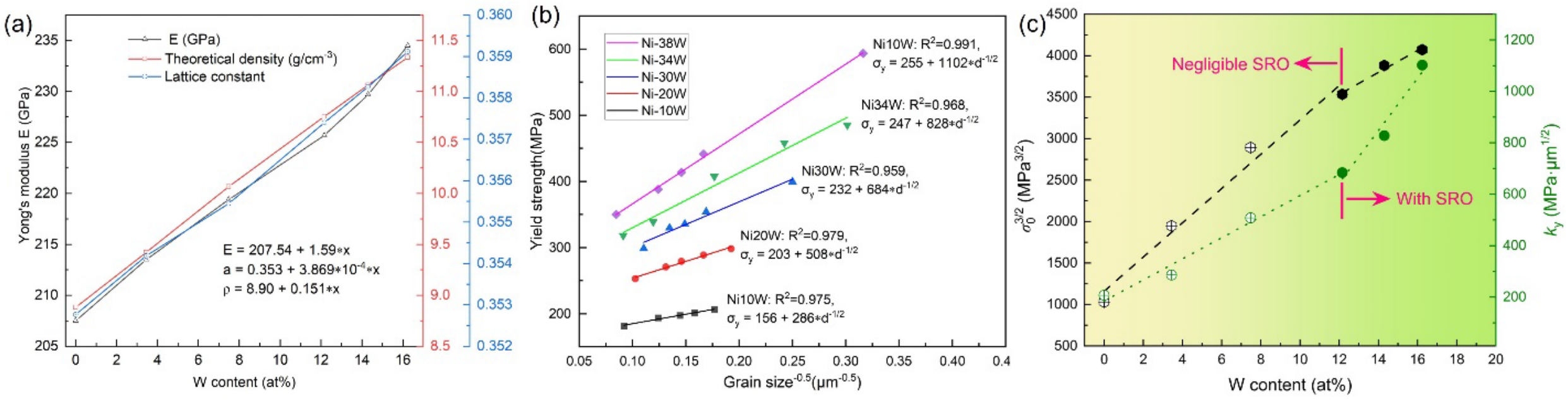

**Figure 3.** (a) Variation of Young's modulus, theoretical density and lattice constant for Ni-W binary alloys as W content increases. (b) Hall-Petch relationships of Ni-W binary alloys with different W contents. The fitted curves for each alloy are indicated. (c) Variation of original yield stress $\sigma_0^{3/2}$ and Hall-Petch constants $k_y$ as W content increases.

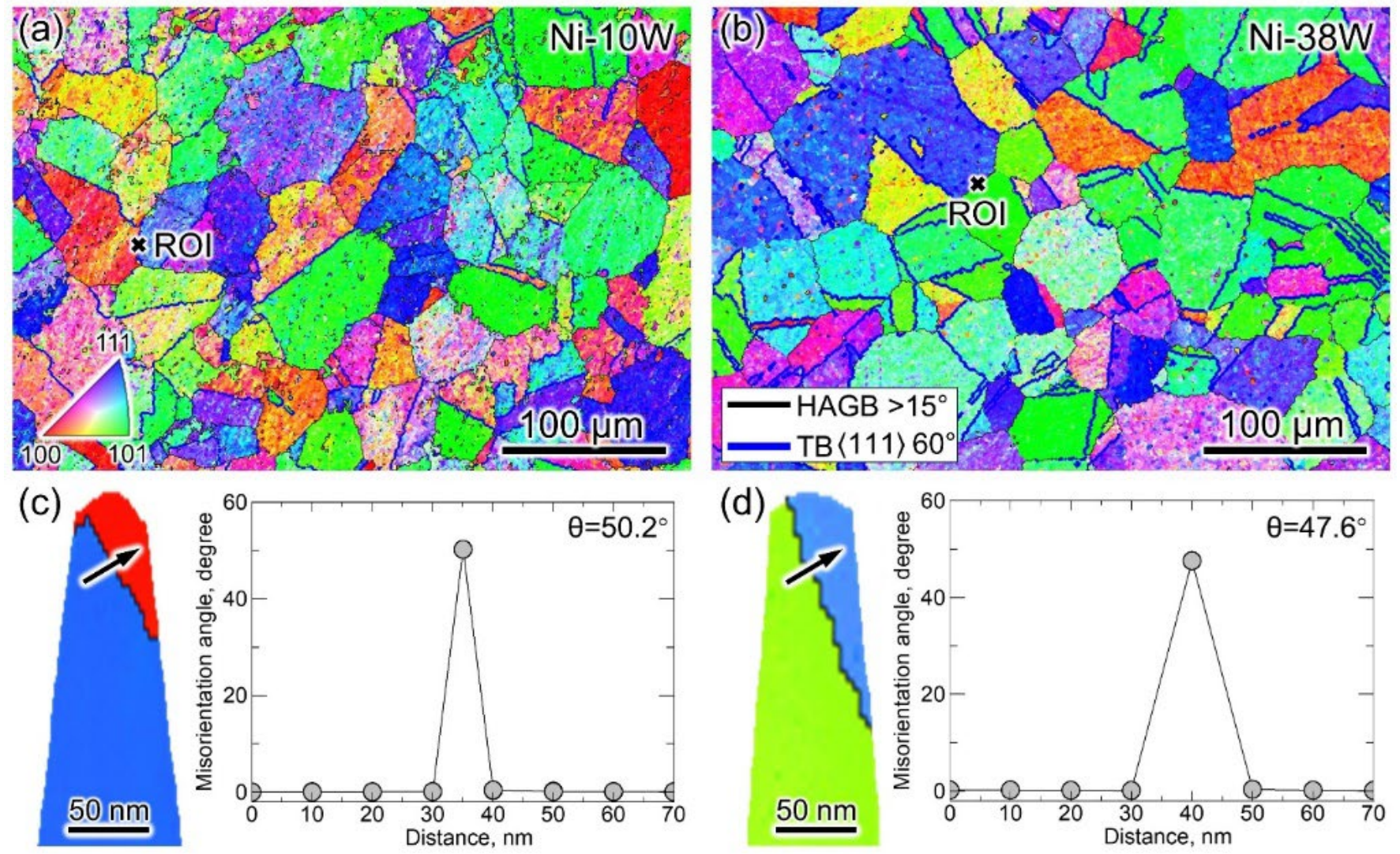

**Figure 4.** (a, b) EBSD inverse pole figure (IPF) maps showing the grain boundaries (GBs) selected for site-specific APT specimen preparation, indicated by region of interest (ROI) cross marks; and (c, d) TKD IPF maps along with corresponding misorientation angle profiles across the GBs indicated by arrows, for Ni-10W and Ni-38W samples solution-treated at 1000 °C and 1100 °C, respectively.

### 3.3 Microstructural features of Ni–W alloys

To investigate the influence of W concentration on GB chemistry, APT analyses were conducted on Ni-10W (solution-treated at 1000 °C) and Ni-38W samples (solution-treated at 1100 °C) alloys with comparable grain sizes. **Figs. 4a** and **4b** show the corresponding IPF maps, and site-specific APT needles were extracted from the regions marked as ROI. Correlative TKD-APT analyses (**Figs. 4c** and **4d**) confirm that the selected GBs in both Ni-10W and Ni-38W possess comparable crystallographic character, each corresponding to a high-angle grain boundary (HAGB) with misorientations of 50° and 48°, respectively. **Fig. 5a** displays the 3D atom maps and 2D ion-density plots of Ni and W for the Ni-10W sample. Although the 3D maps reveal no apparent segregation or depletion, the 2D density plot shows a positional shift of low-density regions associated with crystallographic zone lines (highlighted by black arrows) from the upper to lower regions, indicating the presence of a GB (**Figs. 4a** and **4c**). The 1D composition profile across the GB shows slight W depletion accompanied by Ni enrichment at the GB (**Fig. 5b**). In contrast, as shown in **Figs. 5c** and **5d**, the Ni-38W sample exhibits clear Ni depletion and pronounced W enrichment at a HAGB with the misorientation angle of 48°, indicating a strong segregation tendency at higher W contents. The 1D composition profile quantifies the peak GB W concentration to be ~ 24 at.%. W-depleted zones adjacent to the GB (indicated by arrows) suggests the non-equilibrium segregation state [52].

The slight depletion of W at grain boundaries in Ni–10W suggests a weak or unfavorable thermodynamic driving force for W segregation at this composition, possibly due to competition with residual impurities for favorable boundary sites or variations in local grain-boundary structure. In contrast, the increased chemical potential of W at higher W contents, together with possible co-segregation with impurities, may modify the segregation behavior and promote W enrichment at grain boundaries. Previous atomistic simulations have shown that grain-boundary segregation and chemical ordering in nanocrystalline Ni–W are energetically coupled and strongly composition dependent [53]; W segregation and CSRO are most pronounced at low W concentrations and diminishes with increasing W content. The present APT results, however, reveal an opposite trend, i.e., W is depleted at grain boundaries in the low-W alloy but becomes significantly enriched as the W content increases, accompanied by the formation of CSRO. This discrepancy may arise because the simulations describe an idealized, near-equilibrium binary Ni–W nanocrystalline system, whereas the experimentally observed grain boundaries may be metastable and influenced by processing history, grain-boundary character, residual impurity segregation, and local short-range ordering. Given that the calculated W segregation energy in Ni–W is relatively small and strongly composition dependent, the effective segregation tendency can be readily weakened, modified, or even reversed by these additional factors.

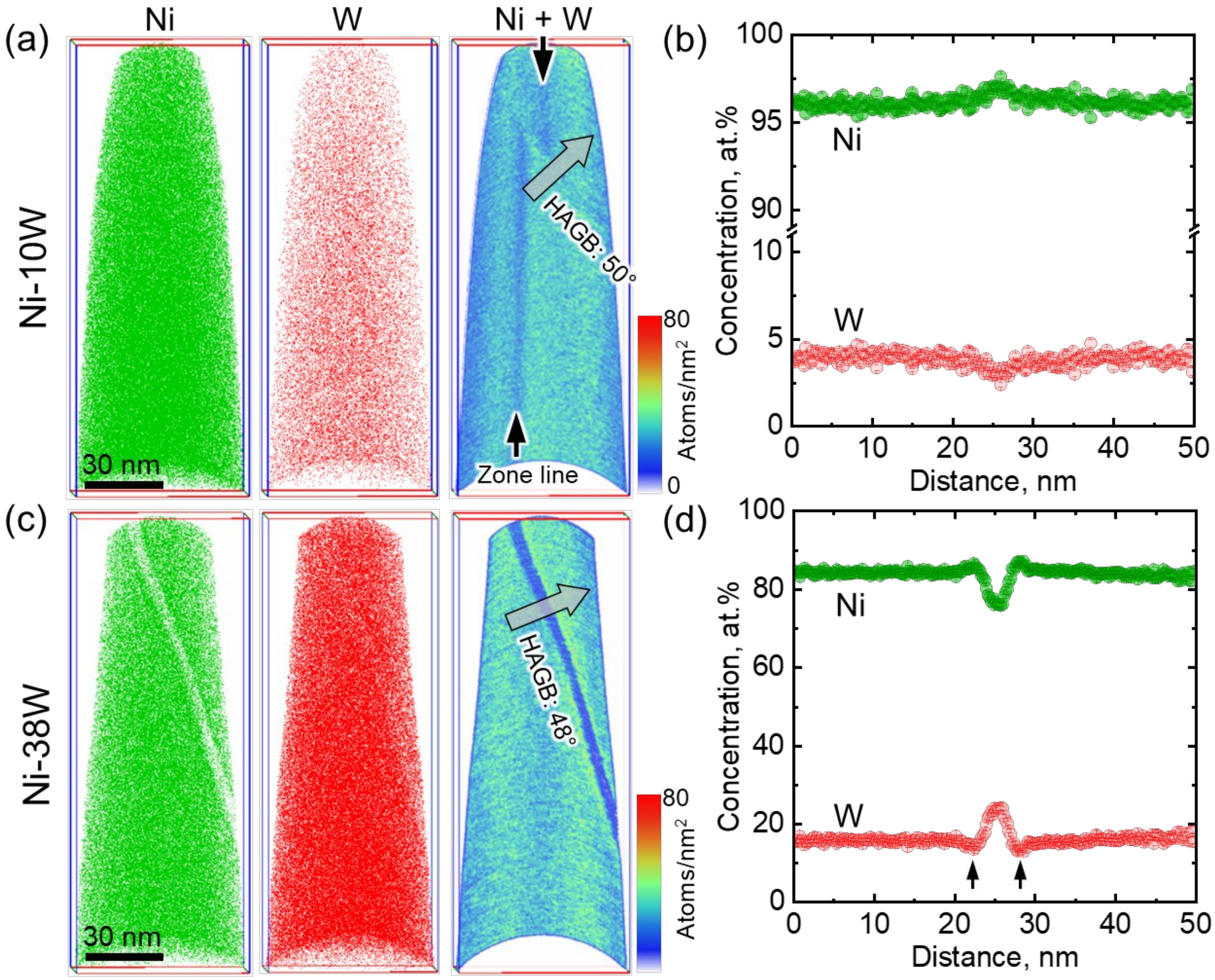

**Figure 5.** (a, c) 3D atom maps and 2D ion density plots of Ni and W; and (b, d) 1D composition profiles across the grain boundaries along the cylindrical regions (30 nm in diameter) indicated by grey arrows in (a) and (c), for Ni-10W and Ni-38W samples solution-treated at 1000 °C and 1100 °C, respectively. Note that the black arrows in (a) indicate low-density regions corresponding to crystallographic zone lines.

To elucidate the effect of W concentration on the deformation microstructures of Ni-W alloys, systematic synchrotron X-ray experiments were conducted on Ni-10W (solution-treated at 1000 °C) and Ni-38W (solution treated at 1100 °C) samples with comparable grain size, subjected to different levels of deformation. The dislocation density was estimated using the classical Williamson–Hall (WH) method, which relates diffraction peak broadening to both grain size and microstrain [54]. Diffraction peaks were fitted using Gaussian functions to determine the full width at half maximum (FWHM), and the strain contribution was analyzed through *K*–*ΔK* plots, where $K = \frac{1}{d} = \frac{2sin\theta}{\lambda} = \frac{Q}{2\pi}$, $\Delta K = \frac{cos\theta * \Delta 2\theta}{\lambda} = \frac{\Delta Q}{2\pi}$ . Here, $\theta$ is the diffraction angle, $Q$ is the reciprocal lattice vector, $\lambda$ is the X-ray wavelength, $d$ is the interplanar spacing, and $\Delta 2\theta$ is the FWHM in real space. If peak broadening arises solely from grain size, the *K*–*ΔK* plot appears as a horizontal line; if dominated by strain, the line passes through the origin. When both effects contribute, the plot intersects the $y$-axis positively. The microstrain (ε) can be further related to dislocation density as $\rho = B(\frac{\varepsilon}{b})^2$, where $B$ is a constant dependent on material properties and dislocation configuration [54]. Importantly, it is noted here that the classical WH method has limitations when applied to anisotropic FCC systems particularly due to elastic anisotropy and contrast-factor effects. However, the simplified analysis was primarily employed to provide a comparative and semi-quantitative evaluation of the relative evolution of dislocation density during deformation, rather than an absolute determination of dislocation density values.

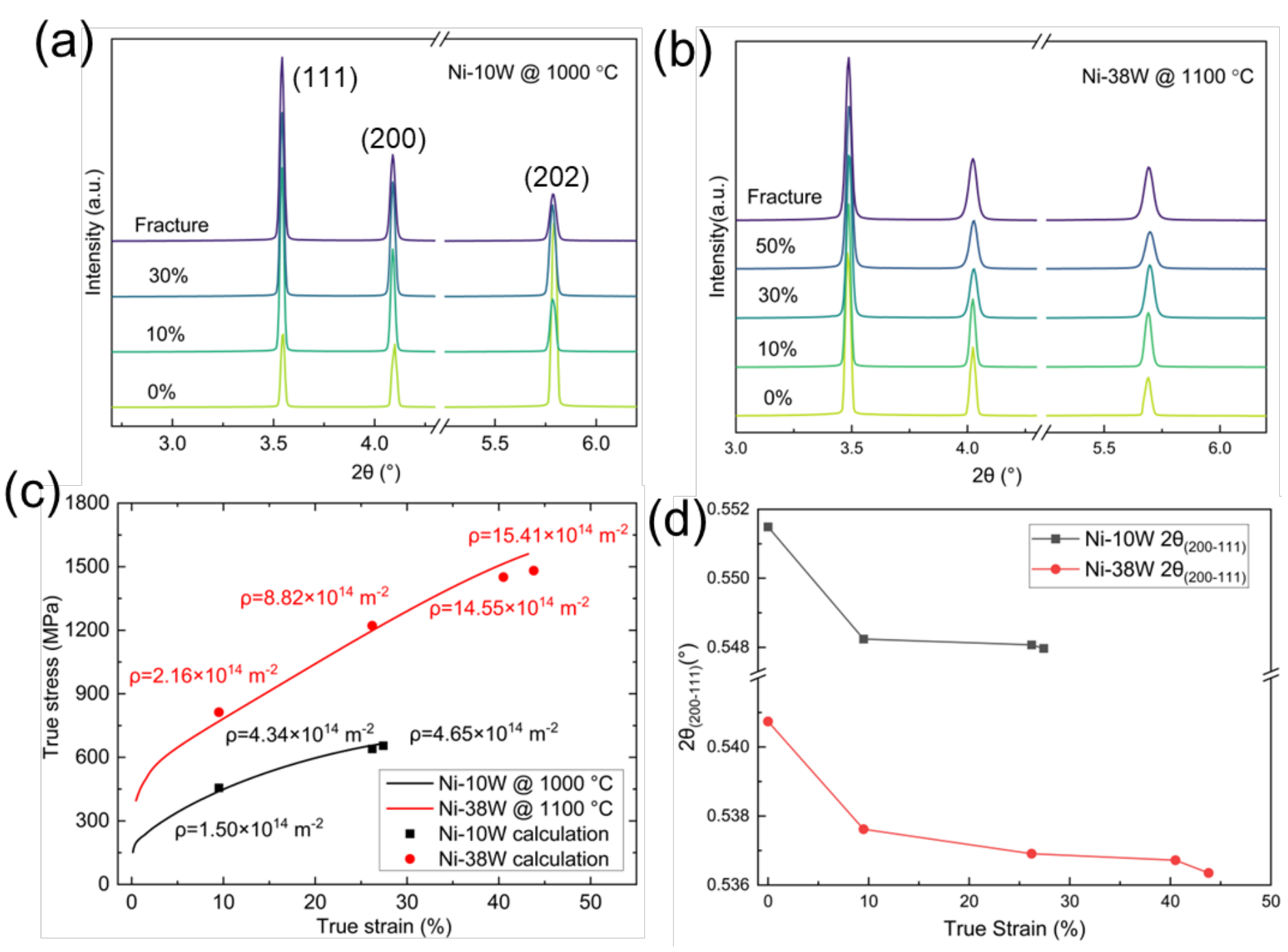

**Figure 6.** Synchrotron radiation XRD of (a) Ni-10W and (b) Ni-38W under various degrees of deformation. (c) True stress–strain curves and corresponding calculated values for Ni-10W and Ni-38W. (d) Comparisons of *2θ* difference between (200) and (111) peak for Ni-10W and Ni-38W alloys. The used photon energy here is 50 keV.

To correct for instrumental broadening, $CeO_2$ powder was used as a standard sample (**Fig. S7a**), as its intrinsic macrostrain is negligible. Gaussian fitting of each diffraction peak — illustrated by the (200) peak in **Fig. S7b**— provided FWHM values used to construct *K*–*ΔK* plots (**Fig. S7c**). The same analysis procedure was applied to Ni-W alloys. **Figs. 6a** and **6b** shows the synchrotron radiation XRD profiles of Ni-10W and Ni3-8W alloys at various deformation levels, with the corresponding *K–ΔK* plots presented in **Figs. S7d** and **7e**. For fractured samples, the true strain corresponds to the maximum true stress, as specimens were taken from regions distant from the fracture surfaces. In both Ni-10W and Ni-38W alloys, the slope of *K–ΔK* plots increased with deformation (**Figs. S7d** and **7e**), indicating a progressive rise in dislocation density. Notably, Ni-38W exhibited a more pronounced slop increase than Ni-10W, signifying enhanced dislocation accumulation. The true stress–strain curves and calculated dislocation densities for both alloys are shown in **Fig. 6c**. Quantitative analysis reveals that Ni-38W possesses significantly higher dislocation densities than Ni-10W — approximately 1.44× and 2.03× greater at 10% and 30% strain, respectively—and this difference further widens to ~3.31× at fracture. The strain-hardening behavior of Ni–W alloys was further examined using the Taylor model, which relates the increment in strain-hardening ( $\sigma_p$ ) to dislocation density ($\rho$) via dislocation interactions:[55] $\sigma_p = M\alpha GbB\rho^{0.5}$, where M is the Taylor factor (~3.06 for FCC metals), $\alpha$ is a fitting constant reflecting the strength of dislocation interactions, *G* is the shear modulus, and *b* is the Burgers vector. Based on a Poisson's ratio of ~0.31, the shear modulus values for Ni-10W (solution-treated at 1000 °C) and Ni-38W (solution-treated at 1100 °C) are ~89.5 GPa and ~81.5 GPa, respectively, and the corresponding Burgers vectors are 0.2504 nm and 0.2539 nm. The fitted $\alpha$ values, which are indicative of dislocations uniformity interactions [56], are ~ 0.338 for Ni-10W and ~0.391 for Ni-38W. This demonstrates that dislocations are more uniform, and dislocation interactions are much stronger in Ni-38W compared to Ni-10W. Consequently, even at comparable dislocation densities, Ni-38W exhibits a larger strengthening increment, underscoring the critical role of dislocation configuration and uniformity in its superior strain-hardening behavior.

Besides dislocations, planar defects such as stacking faults and twins also play important roles in accommodating microscopic deformation. Since planar defects alter the interplanar spacings anisotropically in FCC metals, the (111) reflection—whose plane coincides with the typical fault plane {111}— is particularly to their presence, while the (200) reflection is less affected. Therefore, the *2θ* separation between the (111) and (200) diffraction peaks can serve as a qualitative indicator of the relative stacking fault energy (SFE) [57-59]. To evaluate the influence of W concentrations on the SFE, we analyzed the *2θ* separation between the (111) and (200) peaks of Ni-10W (solution-treated at 1000 °C) and Ni-38W (solution-treated at 1100 °C) to estimate the planar defects probability. As shown in **Figs. 6a** and **6b**, with increasing deformation, the (111) diffraction peak gradually shifts toward higher angles, while the (200) peak remains nearly unchanged. **Fig. 6d**, derived from **Figs. 6a** and **4b**, illustrates the variation in the $2\theta$ difference between the (111) and (200) peaks for Ni-10W and Ni-38W alloys under different deformation levels. At a true strain of 10%, both alloys exhibit a reduced angular difference, indicating the formation of stacking fault. As deformation increases further, the angular difference in Ni-10W becomes nearly constant, suggesting that stacking faults formation saturates beyond 10% strain. In contrast, Ni-38W continues to show a progressive reduction in the $2\theta$ difference, implying continuous accumulation of stacking faults with increasing strain.

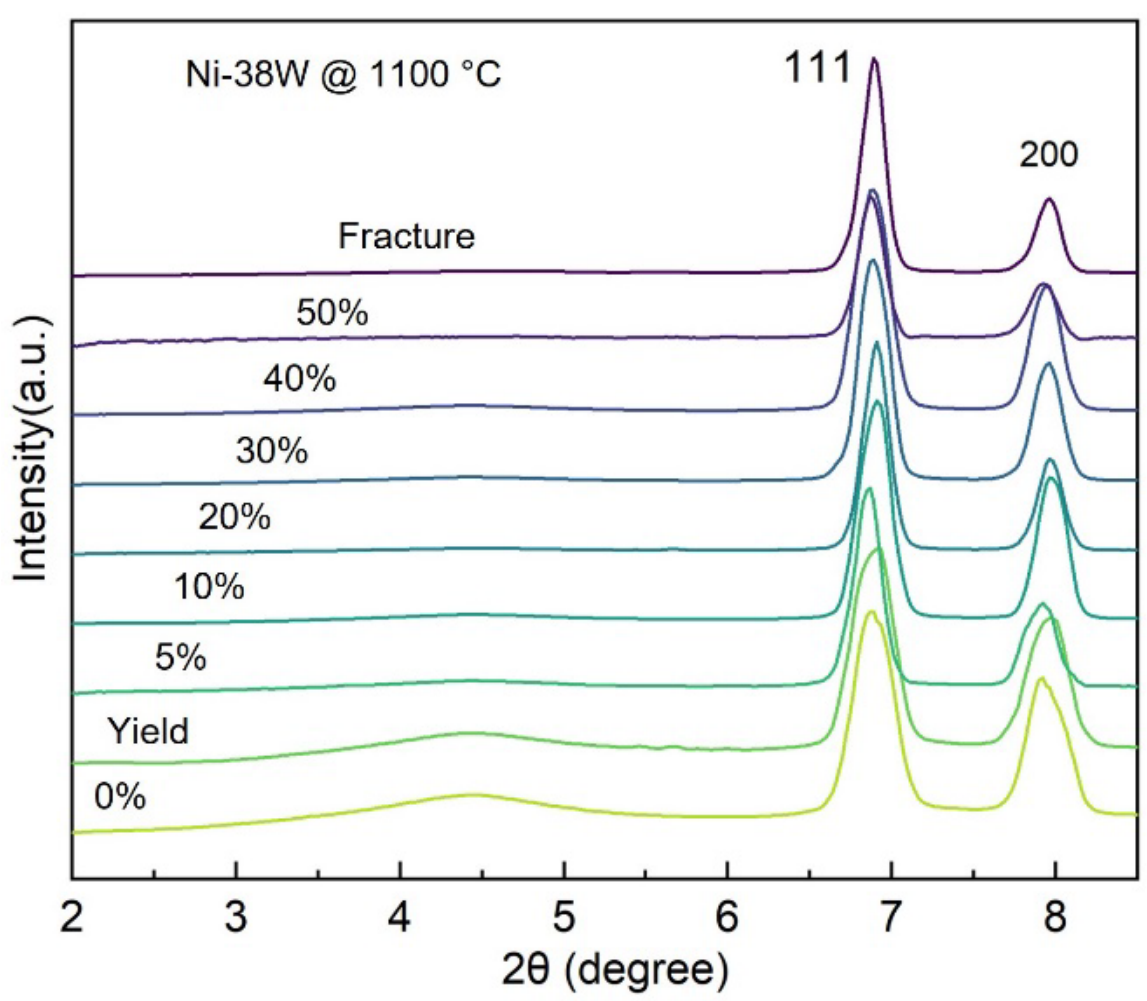

**Figure 7.** Synchrotron radiation XRD of Ni-38W alloy subjected to different strain stages. The used photon energy is 97.9 keV.

**Fig. 7a** corresponds to the synchrotron XRD patterns of Ni-38W at various deformation levels obtained at the low-angle range. It is clearly shown that CSRO features—manifested as a broad peak around 4.5°—disappears after yielding, suggesting that CSRO regions can be destroyed by dislocation shear. To further elucidate the effect of CSRO on the deformation mechanisms with higher spatial resolution, detailed TEM analyses were performed on Ni-10W and Ni-38W alloys subject to varying deformation levels. As shown in **Figs. 8a-d** and Supplementary **Fig. S8**, in Ni-W alloys without negligible CSRO, dislocations are initially activated and progressively accumulated with increasing strain, eventually forming typical dislocation cell structures (highlighted by the dashed frames). In contrast, the Ni-38W alloy exhibiting CSRO features shows pronounced plane slip along {111} planes, leading to the formation of distinct shear bands, as illustrated in **Figs. 8e-g**. Upon further deformation toward fraction, strain localization within these shear bands promotes the development of stacking faults and micro-twins, as observed in **Figs. 8h** and **9**.

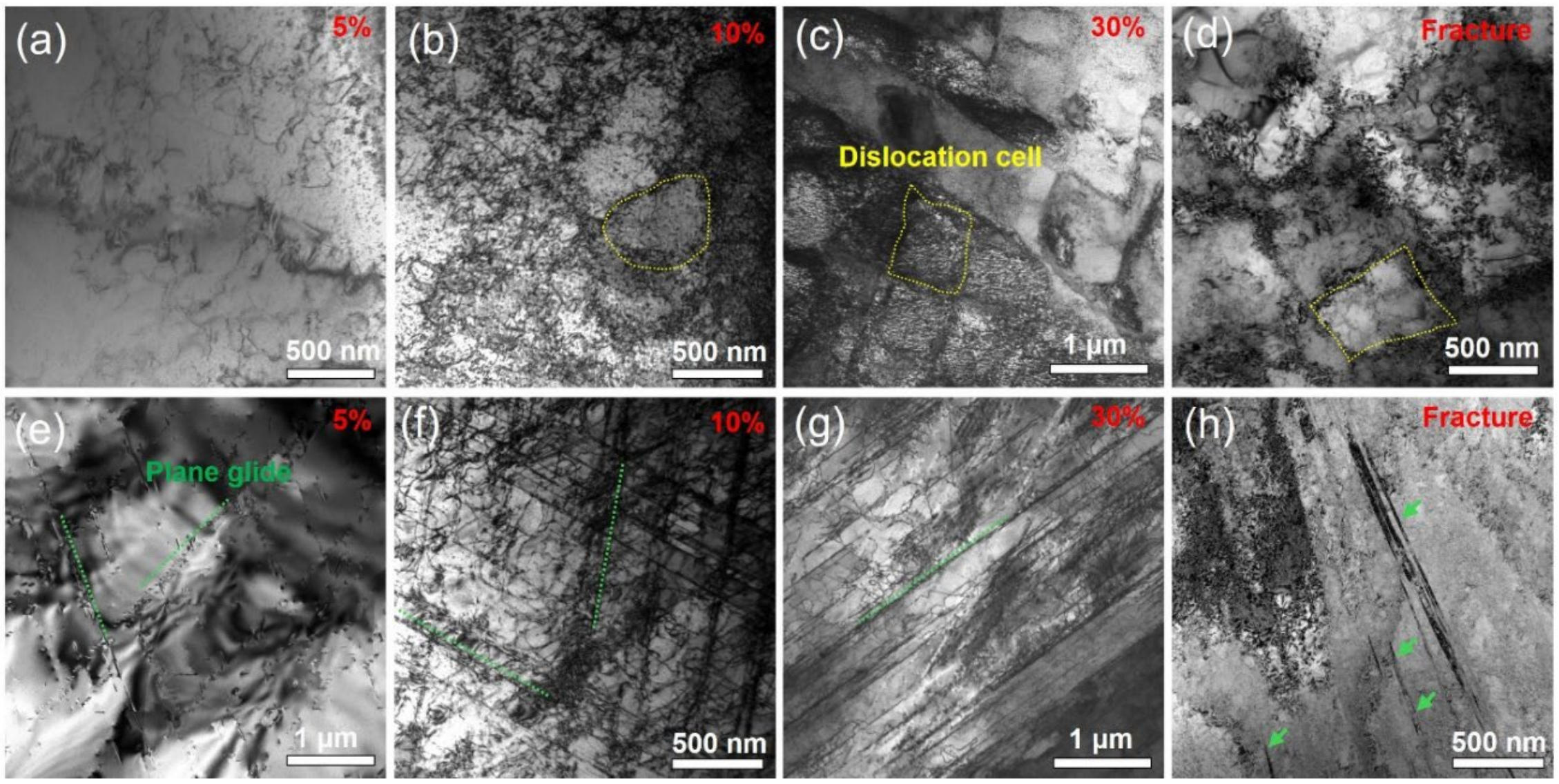

**Figure 8.** TEM image of the (a-d) Ni-10W solution-treated at 1000 °C and (e-h) Ni-38W solution-treated at 1100 °C under different deformation conditions. The dotted yellow frames indicated in (b-d) indicate the formation of dislocation cell structure. The dotted lines in (e-g) indicate the plane glide on {111}. The arrows in (h) indicate the formation of nanosized twin.

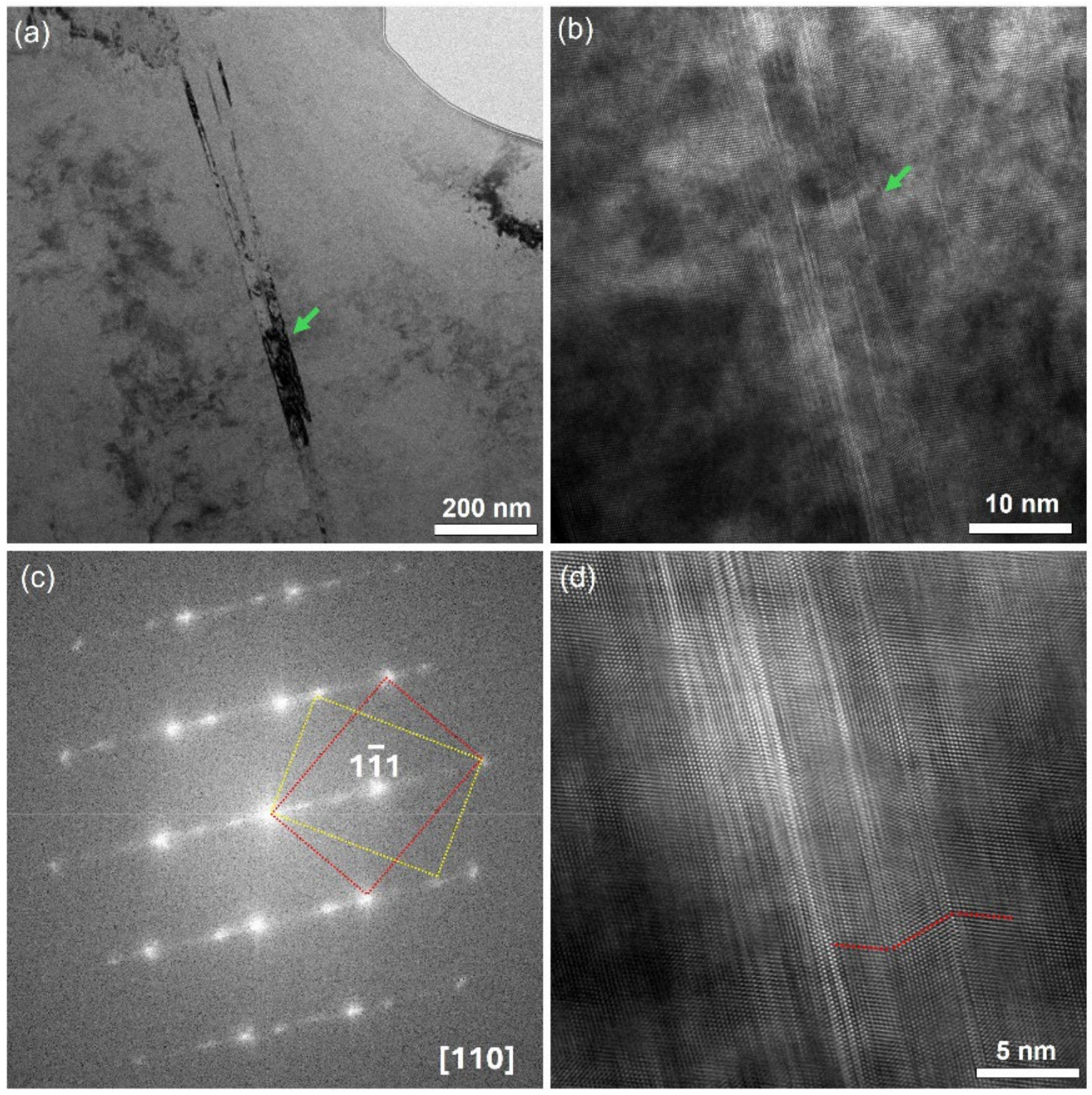

**Figure 9.** (a) TEM image and (b) High resolution TEM image of the Ni-38W alloys solution-treated at 1100 °C under fracture conditions. The arrows indicate the nanosized twin. (c) Digital Fast Fourier transformed patterns corresponding to HRTEM image shown in (b). The dotted frames highlight the twining structure. (d) Localized magnification of HRTEM image shown in (b). The dotted lines indicate the twinning features at the atomic scale.

## 4. Discussion

### 4.1 Origin of CSRO in Ni-W alloys

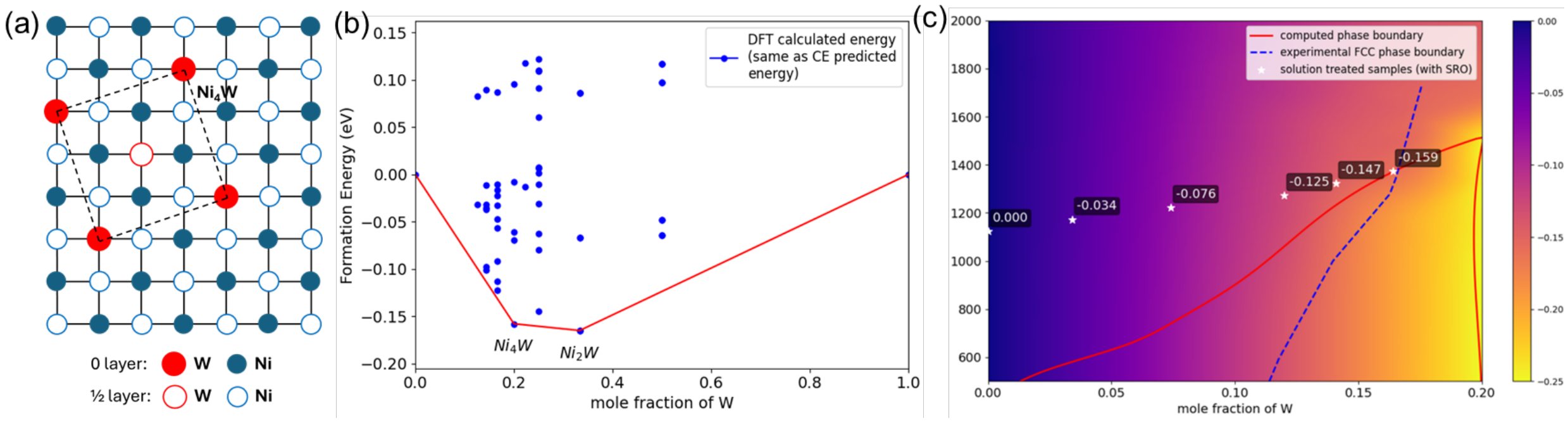

**Figure 10.** (a) Structure of $Ni_4W$ on FCC lattice. (b) The Ni-rich convex hull of FCC Ni-W system (with FCC W as the reference state). (c) The CSRO phase diagram of Ni-rich region compared with experimental FCC phase boundary [60].

The origin of CSRO in the Ni-W binary system can be elucidated through first-principles thermodynamic analysis, as shown in **Fig. 10**. As shown in, the calculated convex hull of the reference structures predicts two ground states (**Fig. 10b**): $D0_{19}$-$Ni_4W$ and $Pt_2Mo$-$Ni_2W$ [61]. Among these, only the $Ni_4W$ phase is relevant under equilibrium conditions, as $Ni_2W$ possesses a higher W concentration and cease to be a global ground state when BCC W is used as reference. The $Ni_4W$ phase exhibits a FCC-based structure, as illustrated in **Fig. 10a**, which is the structural template of the CSRO observed in Ni-W alloys. Based on these reference structures, a CE model was constructed with a cross-valuation (CV) score of 0.014 eV, which is reasonable for the Ni-W system considering its significant lattice mismatch (14.5% in lattice parameters). It is noted that only Ni-rich structures were included in the current CE construction. The discrepancy between the MC-predicted FCC phase boundary and experimental CALPHAD boundary likely originates from the neglect of long-range elastic and vibrational contributions in the present CE formulation [62].

The resulting CSRO phase diagram, shown in **Fig. 10c**, reveals a clear compositional dependence of the CSRO parameters. The detailed procedure for CSRO phase diagram generation is included in the note in Supplementary Materials. As the W content increases from 20 wt.% (7.48 at.%) to 30 wt.% (12.16 at.%), the CSRO parameter for the disordered FCC phase decreases substantially from -0.08 to -0.14, signifying an enhanced degree of CSRO and those exhibiting pronounced local ordering. This CSRO parameter change corroborates with the observed threshold (~30 wt.% or 12.16 at.% W) for significant CSRO formation in our experiments. Such variations in the extent of CSRO have a profound impact on deformation behavior, as evidenced by the corresponding experimental observations (**Figs. 2d-f, 3c**). It is important to note that the present CSRO analysis relies on equilibrium thermodynamics and omits kinetic effects associated with processing, solution treatment, and quenching. Despite this limitation, the analysis is expected to qualitatively reflect the overall trends of CSRO formation in Ni–W alloys.

**4.2 Effect of W concentration and CSRO on friction stress and Hall–Petch coefficient**

As shown in **Fig. 3c**, both $\sigma_0$ and $k_y$ increase with W concentration, reflecting the progressive strengthening of Ni–W alloys through solid-solution and grain boundary effects. However, their concentration dependences differ markedly once CSRO emerges. At low W contents (<30 wt%), the alloy behaves as a conventional FCC solid solution, where strengthening is primarily governed by solute–dislocation interactions arising from atomic size and modulus mismatches.

Consequently, $\sigma_0$ and $k_y$ increase gradually with W addition, consistent with classical solid-solution and grain boundary strengthening mechanisms.

When the W concentration exceeds ~30 wt.%, CSRO becomes pronounced and a clear divergence develops between the evolution of $\sigma_0$ and $k_y$. While the increase in $\sigma_0$ becomes more gradual, $k_y$ exhibits a steep nonlinear rise. The reduced growth rate of $\sigma_0$ can be attributed to the partial replacement of random solute distributions by locally ordered $Ni_4W$-like configurations. Although CSRO increases the resistance to dislocation motion, it simultaneously reduces the randomness of solute–dislocation interactions that dominate conventional solid-solution strengthening. In contrast, CSRO has a much stronger influence on grain-boundary-mediated strengthening. The combination of CSRO and reduced stacking-fault energy promotes planar dislocation glide and suppresses cross-slip, resulting in enhanced dislocation pile-ups at grain boundaries. These pile-ups increase local back-stresses and raise the stress required for slip transmission across grain boundaries, leading to a substantial increase in $k_y$. APT analyses reveal pronounced W enrichment at grain boundaries in the high-W alloys (**Figs. 5b** and **5d**). Previous studies have shown that grain-boundary segregation can further increase the critical stress for dislocation emission and transmission, thereby enhancing Hall–Petch strengthening [63, 64]. The combined effects of planar slip, defect accumulation, and grain-boundary segregation $k_y$, once the CSRO threshold is exceeded.

Particularly noteworthy is the exceptionally high Hall–Petch coefficient obtained for Ni–38W ($\sim 1102\ \text{MPa} \cdot \mu\text{m}^{-\frac{1}{2}}$), which is among the highest reported values for single-phase FCC alloys and exceeds those of most conventional Ni-based alloys and concentrated FCC solid-solution alloys [50, 51]. According to established Hall–Petch models, $k_y$ scales approximately with $(Gb\tau_c)^{1/2}$, where $G$ is the shear modulus, $b$ is the Burgers vector, and $\tau_c$ is the critical resolved shear stress required to activate dislocation sources within grains [65, 66]. W addition substantially increases both the lattice volume misfit and elastic modulus of the Ni matrix, thereby increasing $\tau_c$ and contributing to the high $k_y$ values observed in this system.

However, the increase in $k_y$ cannot be fully explained by conventional solid-solution strengthening. As shown in **Fig. S9**, the experimentally measured $k_y$ increasingly deviates from the theoretical strengthening parameter $G\Delta V^{2/3}/b^2$ when the W concentration exceeds ~30 wt.%, coinciding with the onset of CSRO. This deviation indicates an additional strengthening contribution beyond classical lattice-distortion effects. Together with the observed transition in deformation mode and the emergence of grain-boundary W segregation, these results suggest that CSRO plays a critical role in enhancing defect–grain-boundary interactions, thereby producing the unusually large Hall–Petch strengthening observed in high-W Ni–W alloys.

### 4.3 Correlation between microstructure and deformation behavior

The present study provides direct experimental evidence linking CSRO to the deformation behavior of Ni–W alloys. The mechanical response and corresponding microstructural evolution reveal that CSRO exerts a pronounced influence on dislocation activity, strain hardening behavior, and defect formation mechanisms. These findings elucidate the fundamental role of local chemical ordering in governing the deformation pathways of concentrated solid-solution alloys. At the early stage of deformation, the presence of CSRO in the Ni–38W alloy impedes the motion of dislocations by locally pinning them, thereby elevating the initial yield stress. This behavior is consistent with the higher critical resolved shear stress often observed in CSRO-containing solid solutions. The strain-hardening rate versus strain curves exhibits four distinct stages, among which a characteristic plateau is particularly evident in the CSRO-containing alloy. However, synchrotron XRD data indicate that CSRO features rapidly diminish after yielding (**Fig. 7**), suggesting that the CSRO domains are disrupted by intense dislocation glide and local lattice shearing. Therefore, the observed plateau should not be attributed solely to the sustained presence of CSRO, but rather to the strong dislocation–dislocation interactions and the subsequent planar slip that emerge after the initial breakdown of CSRO.

The distinct differences in deformation microstructures between Ni–10W and Ni–38W provide further insight into the role of W concentration and CSRO in controlling deformation modes. In the Ni–10W alloy, which exhibits relatively weak CSRO and higher SFE, deformation proceeds primarily through the accumulation and entanglement of dislocations, leading to the formation of typical dislocation cell structures (**Figs. 8a-d** and **S8**). In contrast, the Ni–38W alloy, characterized by pronounced CSRO and relatively lower SFE, displays a transition from dislocation activation to planar slip along {111} planes, followed by the development of shear bands, stacking faults, and nanotwins at higher strains (**Figs. 8e-h** and **Fig. 9**). The emergence of planar glide in Ni–38W can be rationalized by the presence of CSRO, which promotes localized slip along specific crystallographic planes by restricting cross-slip and homogenized dislocation motion. Once planar glide is activated, the dislocation arrays on parallel planes strongly interact, giving rise to dense shear bands and facilitating the nucleation of stacking faults and twins. This sequential evolution aligns well with the distinct strain-hardening stages, where the initial strengthening is dominated by CSRO-related lattice friction, the intermediate plateau corresponds to strong planar dislocation interactions, and the later stage is governed by twin-mediated hardening before the onset of softening. The disappearance of CSRO during deformation can be attributed to the cumulative effect of dislocation movement and localized shear strain. The passage of dislocations redistributes solute atoms and disrupts the energetically favorable short-range configurations, thereby destroying the local order that initially impeded slip. Once CSRO is eliminated, deformation proceeds via mechanisms characteristic of random solid solutions, including dynamic recovery and dislocation rearrangement. This transformation from CSRO-controlled to dislocation-interaction–controlled deformation provides a natural explanation for the transition in strain-hardening behavior.

## 5. Conclusions

In summary, this study reveals that the formation of $Ni_4W$-like CSRO in Ni–W alloys is strongly dependent on W concentration, with a threshold near ~30 wt%. This threshold corroborates with the Ni-W CSRO phase diagram predicted by first-principles thermodynamics. Below this threshold, alloys have low CSRO and deform predominantly through dislocation cell structures. As W content increases, the SFE decreases, promoting CSRO formation, and transitioning the deformation mode toward planar glide accompanied by stacking faults and deformation twins. CSRO also exerts a pronounced influence on the Hall–Petch relationship. Beyond the CSRO threshold, enhanced defect–grain boundary interactions lead to a steep, nonlinear rise in the Hall–Petch slope ($k_y$), whereas the increase in friction stress ($\sigma_0$) becomes more gradual. This contrast indicates that CSRO more effectively strengthens grain boundary–related mechanisms than intrinsic lattice resistance. Additionally, Ni–W alloys containing CSRO exhibits a characteristic plateau occupying most of the deformation stage in the strain-hardening rate versus strain curve, reflecting a sustained balance between hardening and softening throughout deformation. Overall, W concentration and the associated development of CSRO play a pivotal role in governing the deformation behavior and mechanical stability of Ni–W alloys. These insights provide valuable guidance for the compositional and microstructural design of next-generation Ni–W–based MHAs, enabling optimized combinations of density, strength, ductility, and toughness for advanced structural applications.

## Acknowledgments

We acknowledge the Shanghai Synchrotron Radiation Facility for the X-ray diffraction experiments. This work also made use of the EPIC facility (RRID: SCR_026361) of Northwestern University's NU*ANCE* Center, which has received support from the SHyNE Resource (NSF ECCS-2025633), the IIN, and Northwestern's MRSEC program (NSF DMR-2308691). Hantong and Bi-Cheng acknowledge the funding support from NSF through the CAREER Grant DMR-2042284. Atom probe tomography was performed at the Northwestern University Center for Atom-Probe Tomography (NUCAPT, RRID: SCR_017770). NUCAPT received support from the MRSEC program (NSF DMR-2308691) at the Materials Research Center and the SHyNE Resource (NSF ECCS-2025633) at Northwestern University.

# Supporting Materials

## Concentration-Dependent Tungsten Effects on Chemical Short-Range Order and Deformation Behavior in Ni–W alloys

Shaozun Liu[1,6*], Zehao Li[2], Hantong Chen[3], Xingyuan San[4], Bi-Cheng Zhou[3*], Dieter Isheim[2,7], Tiejun Wang[1], Nie Zhao[5], Yu Liu[1], Yong Gan[1], Xiaobing Hu[2,8*]

[1]Central Iron and Steel Research Institute, Beijing 100081, China

[2]Department of Materials Science and Engineering, Northwestern University, Evanston, IL 60208, USA

[3]Department of Materials Science and Engineering, University of Virginia, Charlottesville, Virginia 22904, USA

[4]Hebei Key Lab of Optic-electronic Information and Materials, The College of Physics Science and Technology, Hebei University, Baoding 071002, China

[5]School of Materials Science and Engineering, Xiangtan University, Xiangtan 411105, China

[6]Yangjiang Advanced Alloys Laboratory, Yangjiang 529500, China

[7]Northwestern University Center for Atom Probe Tomography (NUCAPT), Evanston, IL 60208, USA

[8]The NU*ANCE* Center, Northwestern University, Evanston, IL 60208, USA

[*]Corresponding authors. Email: xbhu@northwestern.edu (X.H.); liushaozun@cisri.com (S.L.); bicheng.zhou@virginia.edu (B.C.Z.);

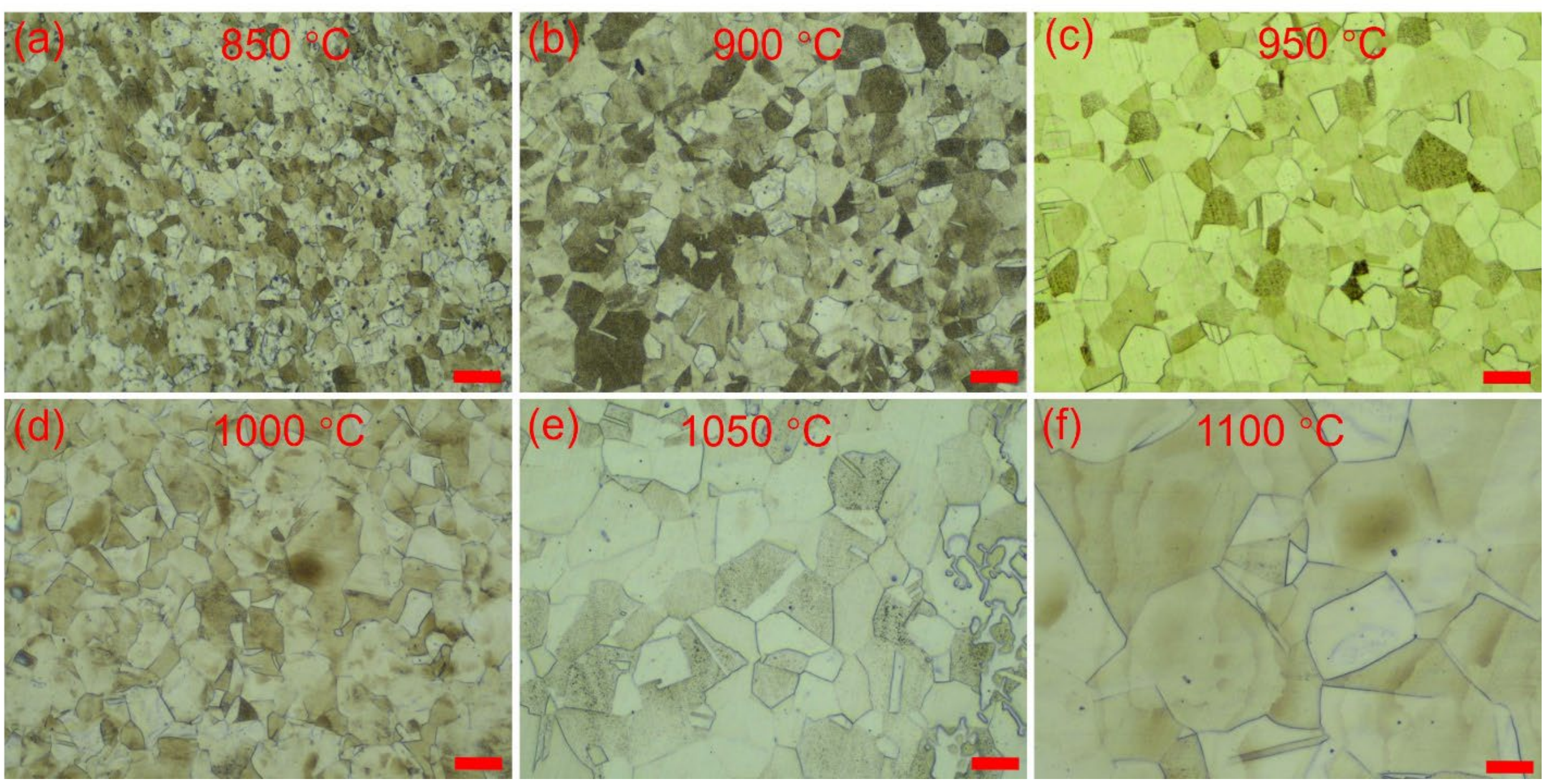

**Figure S1.** Optical micrograph of forged Ni-10W sample subjected to different solution temperatures. The scar bars in (a-f) are 50 μm.

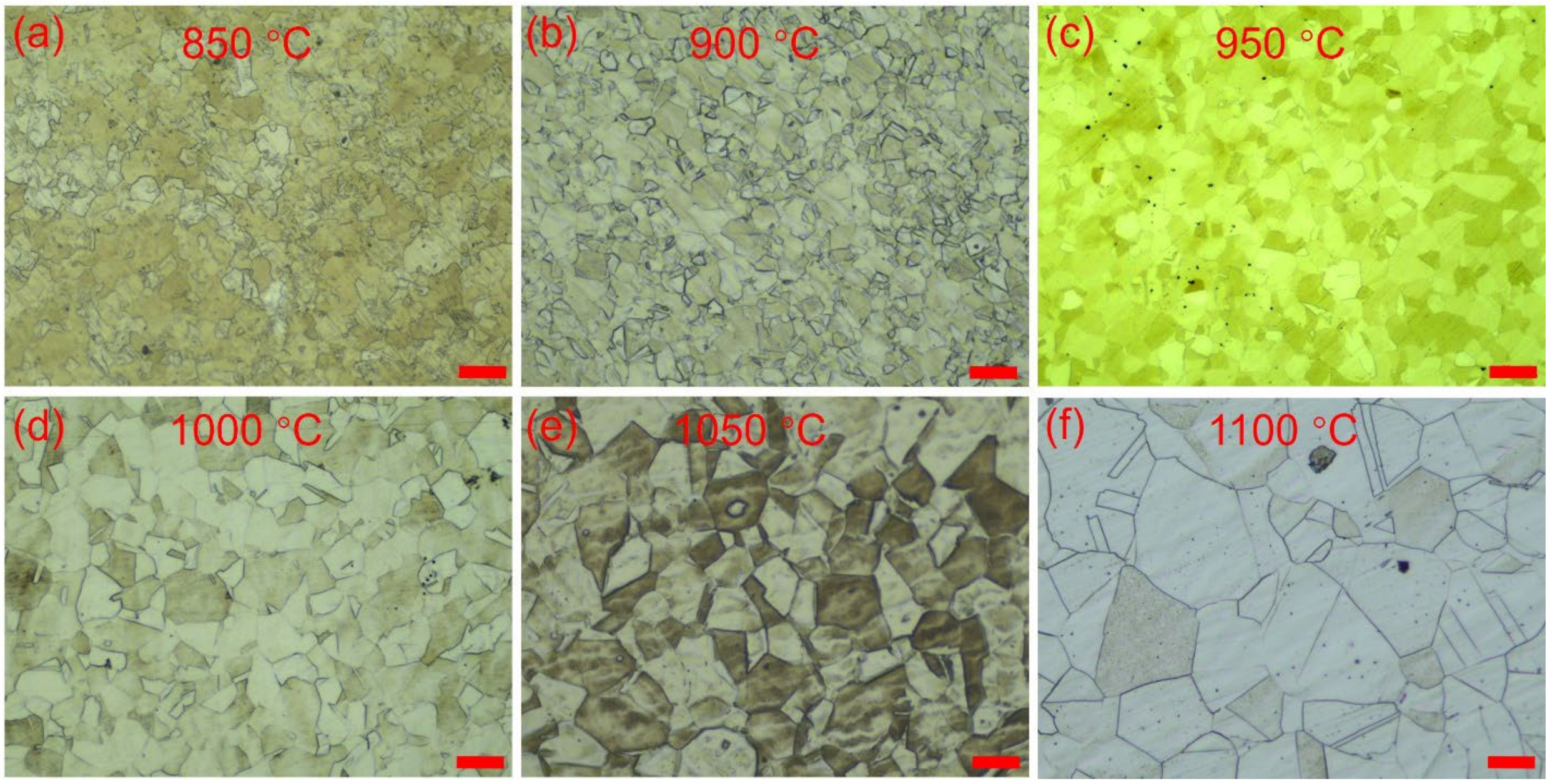

**Figure S2.** Optical micrograph of forged Ni-20W sample subjected to different solution temperatures. The scar bars in (a-f) are 50 μm.

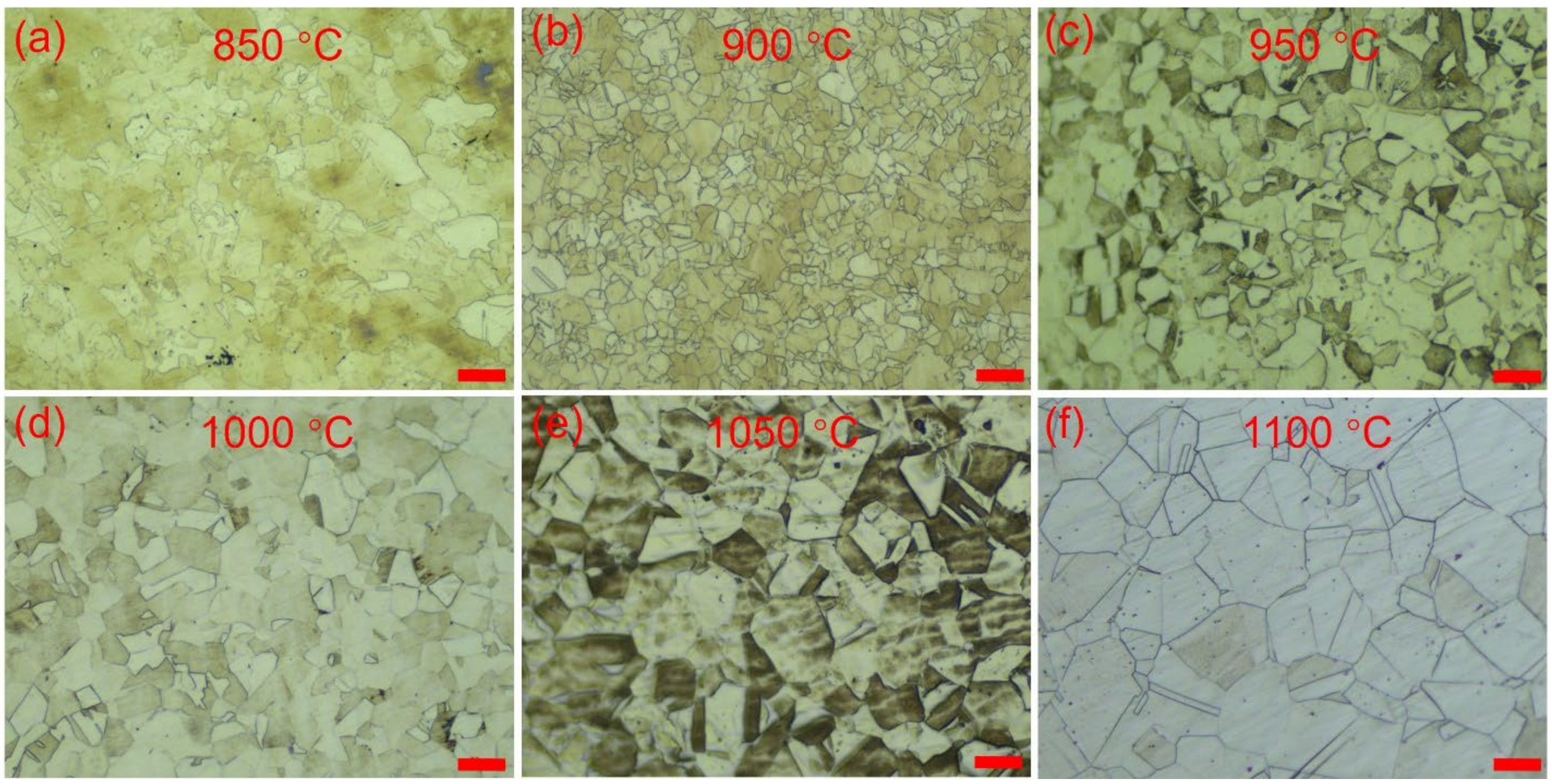

**Figure S3.** Optical micrograph of forged Ni-30W sample subjected to different solution temperatures. The scar bars in (a-f) are 50 μm.

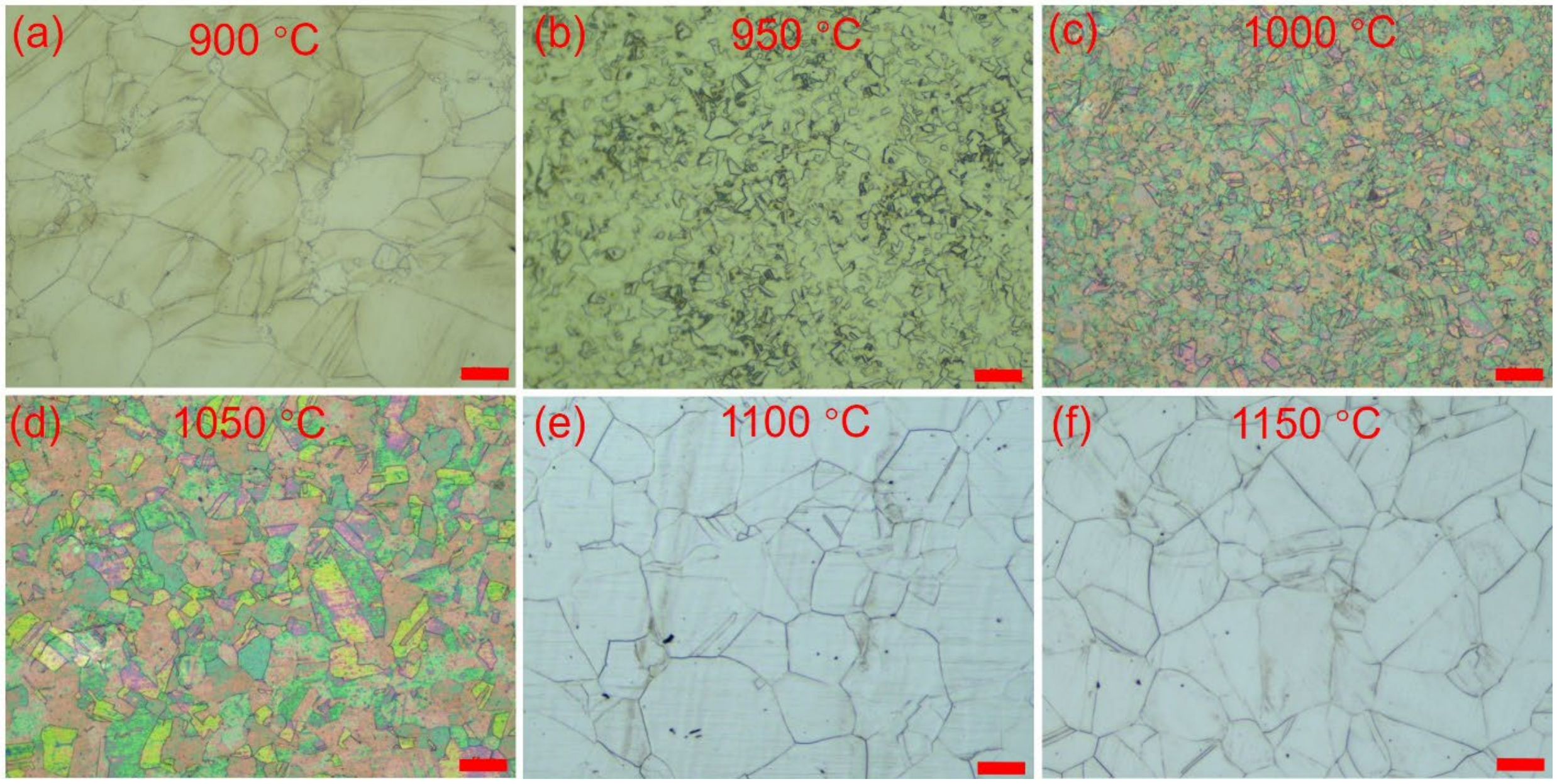

**Figure S4.** Optical micrograph of forged Ni-34W sample subjected to different solution temperatures. The scar bars in (a-f) are 50 μm. The abnormal microstructures in (a) indicates that the 900 °C is still not high enough to initiate the re-crystallization process for forged Ni-34W alloy.

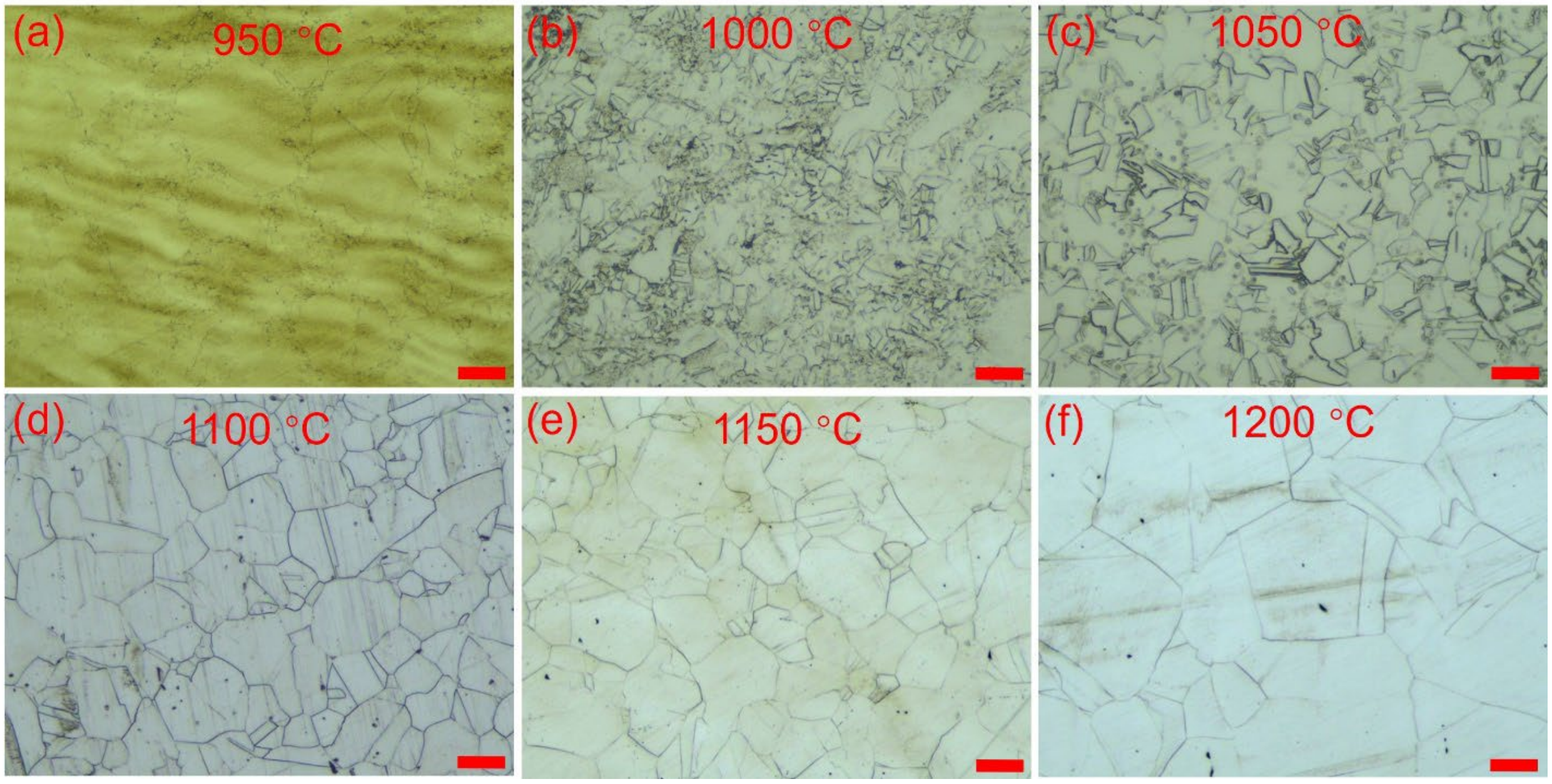

**Figure S5.** Optical micrograph of forged Ni-38W sample subjected to different solution temperatures. The scar bars in (a-f) are 50 μm. The abnormal microstructures in (a) indicate that the 950 °C is still not high enough to initiate the re-crystallization process for forged Ni-38W alloy.

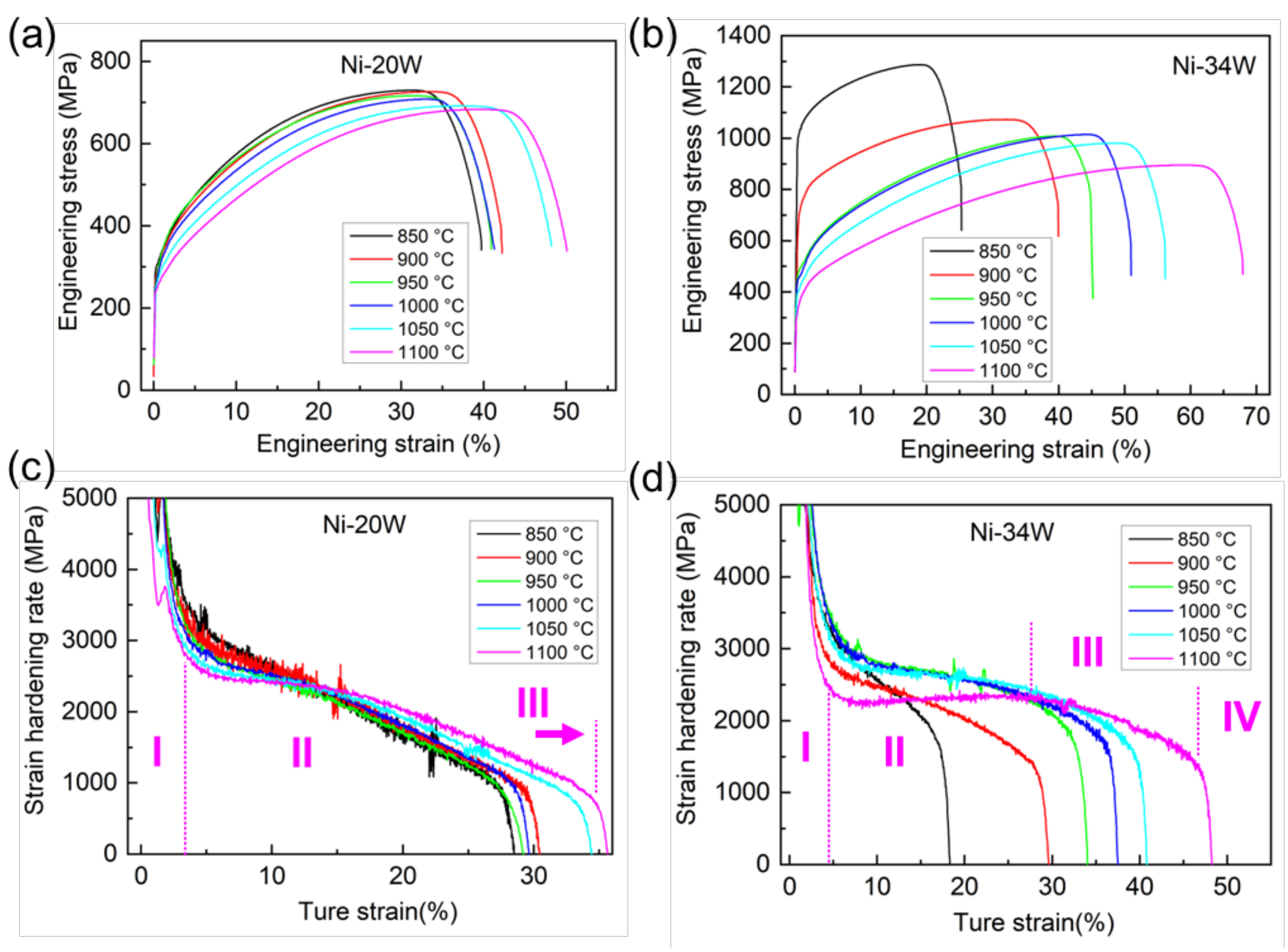

**Figure S6.** Engineering stress-strain and strain hardening rate curves of Ni-W binary alloys at different solution temperatures. (a, c) Ni-20W, (b, d) Ni-34W. The solution temperature for each curve is indicated in the legend. The vertical pink dash lines in (c) and (d) highlight the distinct deformation stages for Ni-20W solution-treated at 1100 °C, and Ni-34W solution-treated at 1100 °C respectively.

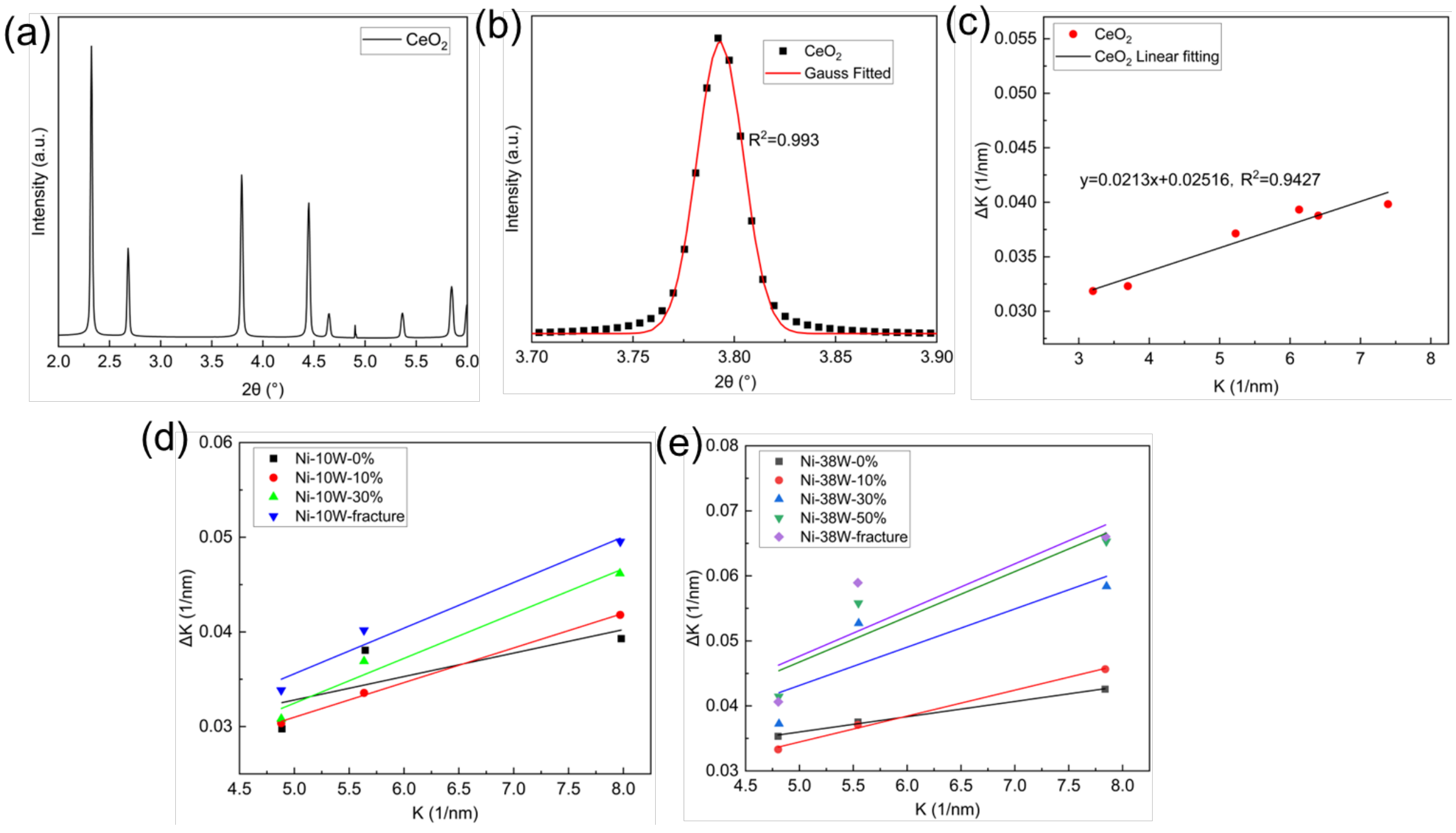

**Figure S7.** (a-c) Diffraction spectra, corresponding fitting results, and *K*–*ΔK* plots for $CeO_2$. *K*–*ΔK* plots of (d) Ni-10W and (e) Ni-38W under various degrees of deformation. Ni-10W and Ni-38W are solution-treated at 1000 °C and 1100 °C, respectively.

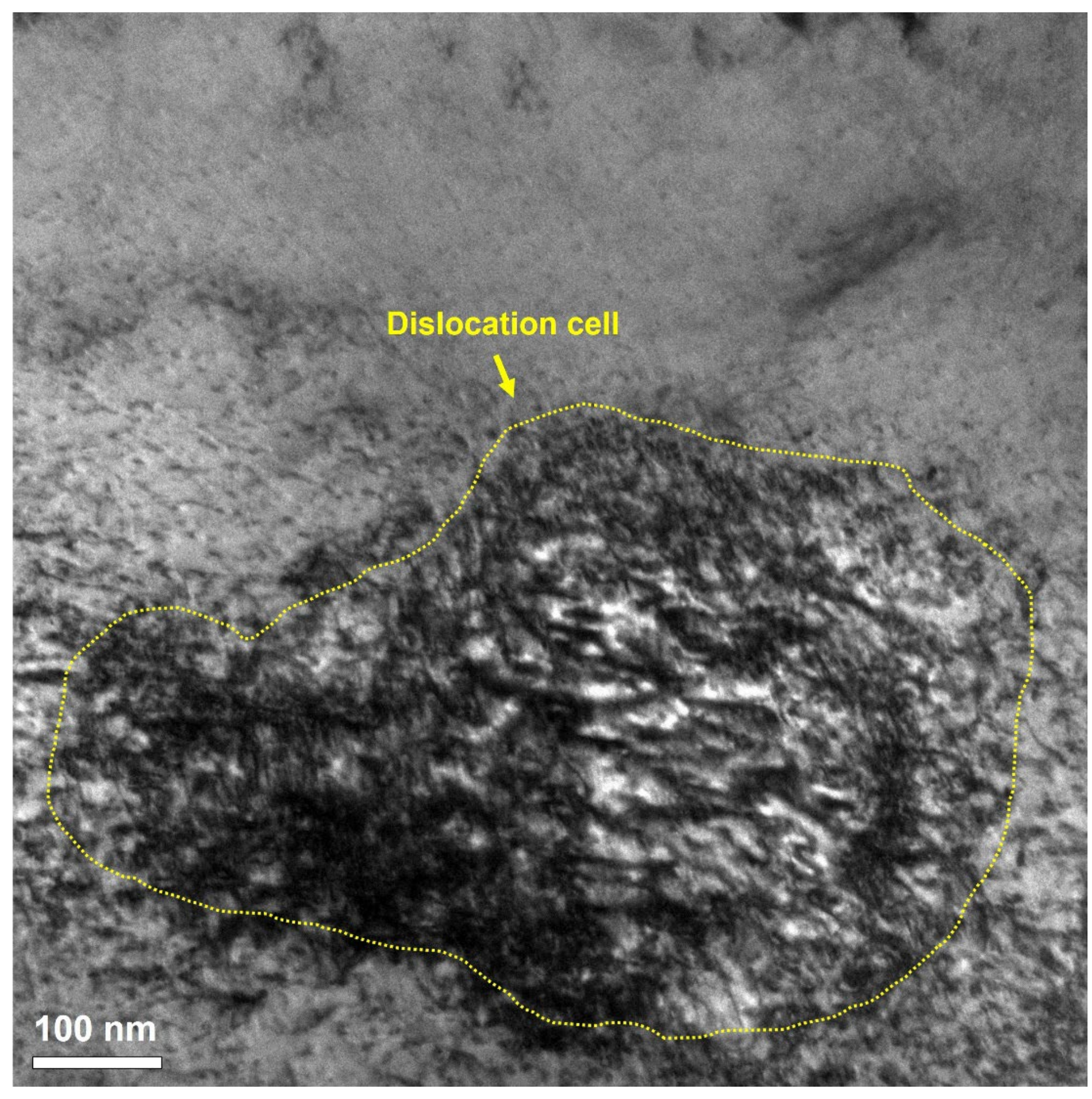

**Figure S8.** High magnification TEM image showing a typical dislocation cell structure within Ni-10W alloys (solution-treated at 1000 °C) deformed at 30%.

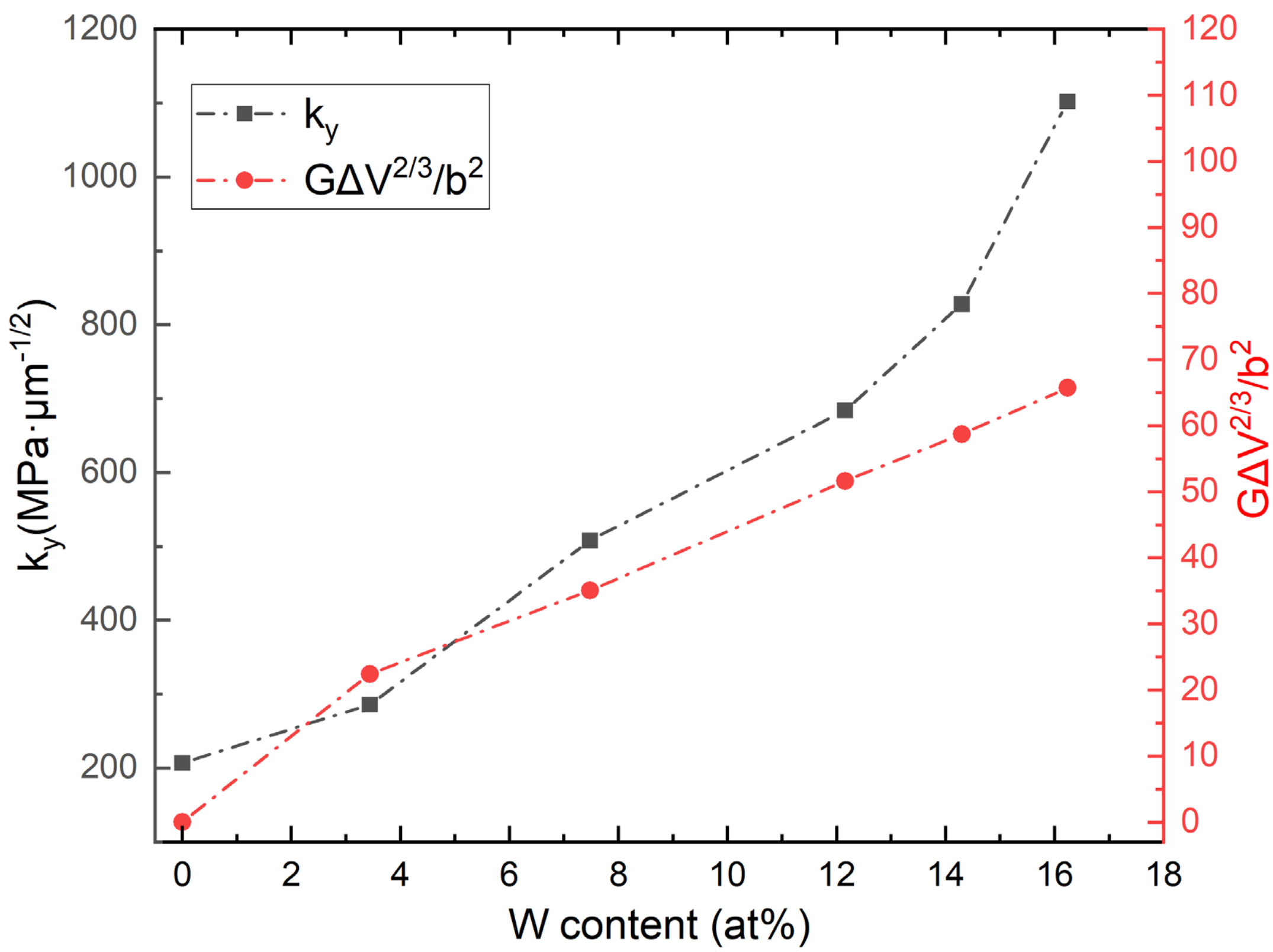

**Figure S9.** Variation of $k_y$ and the theoretical strengthening parameter $G\Delta V^{2/3}/b^2$ with W atomic concentration in Ni–W alloys.

**Table S1**. Grain size (μm) of Ni-W binary alloys with varied W contents (wt.%) and solution temperatures (°C)

| Solution temperature | Ni-10W | Ni-20W | Ni-30W | Ni-34W | Ni-38W |
|---|---|---|---|---|---|
| 900 | 32 | 27 | 16 | NA | NA |
| 950 | 40 | 36 | 35 | 11 | NA |
| 1000 | 48 | 47 | 45 | 17 | 10.5 |
| 1050 | 65 | 58 | 55 | 32 | 36 |
| 1100 | 118 | 95 | 82 | 70 | 47 |
| 1150 | NA | NA | NA | 120 | 65 |
| 1200 | NA | NA | NA | NA | 140 |

Note: “NA” indicates “not appliable” for the specific alloys due to either too high or too low solution temperatures.

**Table S2**. Yield strength (MPa) of Ni-W binary alloys with varied W contents (wt.%) and solution temperatures (°C)

| Solution temperature | Ni-10W | Ni-20W | Ni-30W | Ni-34W | Ni-38W |
|---|---|---|---|---|---|
| 900 | 206 | 298 | 399 | 724 | NA |
| 950 | 201 | 289 | 354 | 485 | 1290 |
| 1000 | 198 | 279 | 336 | 458 | 593 |
| 1050 | 194 | 271 | 330 | 407 | 441 |
| 1100 | 181 | 253 | 299 | 339 | 414 |
| 1150 | NA | NA | NA | 318 | 388 |
| 1200 | NA | NA | NA | NA | 350 |

Note: “NA” indicates “not appliable” for the specific alloys due to either too high or too low solution temperatures.

**Table S3**. Ultimate tensile strength (MPa) of Ni-W binary alloys with varied W contents (wt.%) and solution temperatures (°C)

| Solution temperature | Ni-10W | Ni-20W | Ni-30W | Ni-34W | Ni-38W |
| --- | --- | --- | --- | --- | --- |
| 900 | 525 | 730 | 930 | 1077 | NA |
| 950 | 517 | 719 | 900 | 1009 | 1598 |
| 1000 | 514 | 710 | 854 | 1017 | 1146 |
| 1050 | 508 | 693 | 846 | 984 | 1026 |
| 1100 | 506 | 685 | 819 | 897 | 1015 |
| 1150 | NA | NA | NA | 827 | 938 |
| 1200 | NA | NA | NA | NA | 874 |

Note: "NA" indicates "not appliable" for the specific alloys due to either too high or too low solution temperatures.

**Table S4**. Elongation (%) of Ni-W binary alloys with varied W contents (wt.%) and solution temperatures (°C)

| Solution temperature | Ni-10W | Ni-20W | Ni-30W | Ni-34W | Ni-38W |
| --- | --- | --- | --- | --- | --- |
| 900 | 45 | 42 | 48.5 | 39.50 | NA |
| 950 | 45 | 40.5 | 52.5 | 44.5 | 20.0 |
| 1000 | 44 | 41 | 56 | 50.5 | 50.5 |
| 1050 | 42.5 | 48 | 58.5 | 55.5 | 53 |
| 1100 | 43.5 | 48 | 60 | 67 | 58 |
| 1150 | NA | NA | NA | 66 | 63 |
| 1200 | NA | NA | NA | NA | 66.5 |

Note: "NA" indicates "not appliable" for the specific alloys due to either too high or too low solution temperatures.

**Note: Detailed procedure for SRO phase diagram generation**

To start, cluster expansion was performed using the *mmaps* code in ATAT. The composition range was limited to 50 at% W, excluding the end member. The reference energy of FCC W was determined using the inflection detection method without lattice relaxation to account for its mechanical instability. The Bayesian fitting scheme was employed to obtain effective cluster interactions (ECIs) using the command:

*mmaps -fa=bayesian*

Details regarding the cluster expansion (CE) are documented in the following Github repository: https://github.com/cht265358/Ni-W-CE-calculation-files. The repository provides the ECI files and clusters used in CE construction, as well as fitting results and ground state information.

After obtaining the ECI, the *phb* code was used to determine the phase boundary between $Ni_4W$ and Ni. A representative command is:

*phb -keV -er=15 -gs1=1 -gs2=2 -dx=1.0e-3 -T=100 -dT=25 -mu=-0.4*

Here *-gs1=1* and *-gs2=2* specify the equilibrated phases listed in *gs_str.out*, and the chemical potential (*-mu*) should be adjusted to ensure correct phase boundary tracking.

The Monte Carlo simulation was performed using the *memc2* code to evaluate equilibrium configurations:

*memc2 -keV -is=str.out -er=15 -n=100 -eq=500*

Where *-is* denotes, the initial structure provides the starting composition for MC simulation. Additional parameters were defined in the *control.in* file.

The V-matrix, required to convert correlation functions into cluster probabilities, was generated using the *cvmclus* command:

*cvmclus -d*

To run *cvmclus*, a *maxclus.in* file is required. For calculating the SRO parameter for the first nearest neighbor, the *maxclus.in* file is just the smallest binary cluster. The code will generate the V matrix in *vmat.out* while also generating a *configmult.out* file. To obtain the SRO, we also need to extract the correlation matrix from the corresponding columns in the *mc.out* file generated through MC simulation by *memc2*. Then we need to multiply the correlation matrix with the V-matrix to obtain the cluster probability. Finally, the cluster probability is multiplied with the multiplicity in *configmult.out* to obtain the true cluster probability.